\documentclass[twocolumn,times,trackchanges]{aastex63}
\usepackage[utf8]{inputenc}
\usepackage[normalem]{ulem}
\usepackage{amsmath}
\usepackage{booktabs}

\newcommand{\be}{\begin{eqnarray}}
\newcommand{\ee}{\end{eqnarray}}

\newcommand{\Ftilde}{\widetilde{F}}
\newcommand{\DM}{{\rm DM}}

\newcommand{\Gscatt}{G_{\rm scatt}}

\newcommand{\Pne}{P_{\delta n_e}}
\newcommand{\cnsq}{{\rm C_n^2}}
\newcommand{\linner}{l_{\rm i}}
\newcommand{\louter}{l_{\rm o}}

\providecommand{\DMh}{\DM_{\rm h}}

\renewcommand{\Gscatt}{G_{\rm scatt}}
\newcommand{\dsl}{d_{\rm sl}}
\newcommand{\dlo}{d_{\rm lo}}
\newcommand{\dso}{d_{\rm so}}
\newcommand{\thetad}{\theta_{\rm d}}
\newcommand{\nud}{\Delta \nu_{\rm d}}

\turnoffeditone

\begin{document}

\title{Radio Scattering Horizons for Galactic and Extragalactic Transients}
\author[0000-0002-4941-5333]{Stella Koch Ocker}
\affiliation{Department of Astronomy and Cornell Center for Astrophysics and Planetary Science, Cornell University, Ithaca, NY, 14853, USA}

\author[0000-0002-4049-1882]{James M. Cordes}
\affiliation{Department of Astronomy and Cornell Center for Astrophysics and Planetary Science, Cornell University, Ithaca, NY, 14853, USA}

\author[0000-0002-2878-1502]{Shami Chatterjee}
\affiliation{Department of Astronomy and Cornell Center for Astrophysics and Planetary Science, Cornell University, Ithaca, NY, 14853, USA}

\author[0000-0002-3135-3824]{Miranda R. Gorsuch}
\affiliation{Department of Physics and Astronomy, University of Wisconsin Stevens Point, Stevens Point, WI, 54481}

\correspondingauthor{Stella Koch Ocker}
\email{sko36@cornell.edu}

\received{2022 March 30}

\revised{2022 May 27}

\accepted{2022 June 1}

\begin{abstract}
Radio wave scattering can cause severe reductions in detection sensitivity for surveys of Galactic and extragalactic fast ($\sim$ms duration) transients. While Galactic sources like pulsars undergo scattering in the Milky Way interstellar medium (ISM), extragalactic fast radio bursts (FRBs) can also experience scattering in their host galaxies and other galaxies intervening their lines-of-sight. We assess Galactic and extragalactic scattering horizons for fast radio transients using a combination of NE2001 to model the dispersion measure (DM) and scattering time ($\tau$) contributed by the Galactic \edit1{disk}, and independently constructed electron density models for the \edit1{Galactic halo } and other galaxies' ISMs and halos that account for different galaxy morphologies, masses, densities, and strengths of turbulence. \edit1{For source redshifts $0.5\leq z_{\rm s}\leq1$, an all-sky, isotropic FRB population has simulated values of $\tau(1\rm~GHz)$ ranging from $\sim1~\mu$s to $\sim2$ ms (90\% confidence, observer frame) that are dominated by host galaxies, although $\tau$ can be $\gg2$~ms at low Galactic latitudes. A population at $z_{\rm s}=5$ has $0.01\lesssim\tau\lesssim300$~ms at 1~GHz (90\% confidence), dominated by intervening galaxies. About $20\%$ of these high-redshift FRBs are predicted to have $\tau>5$ ms at 1~GHz (observer frame), and $\gtrsim40\%$ of FRBs between $z_{\rm s}\sim0.5-5$ have $\tau\gtrsim1$ ms for $\nu\leq 800$ MHz. Our scattering predictions may be conservative if scattering from circumsource environments is significant, which is possible under specific conditions. The percentage of FRBs selected against from scattering could also be substantially larger than we predict if circumgalactic turbulence causes more small-scale ($\ll1$~au) density fluctuations than observed from nearby halos.}
\end{abstract}

\section{Introduction}

The known population of astrophysical radio transient sources has expanded dramatically over the past decade with the advent of widefield radio surveys targeting both pulsars and fast radio bursts (FRBs). Such surveys underscore the utility of these sources as versatile probes of astrophysical phenomena, ranging from the nature of compact objects \citep[e.g.][]{2016ARAA..54..401O} to the distribution of plasma within the Milky Way and across the intergalactic medium (IGM; e.g. \citealt{2020Natur.581..391M}), and even to the cumulative background of gravitational waves produced by supermassive black hole binaries across the Universe \citep[e.g.][]{2010CQGra..27h4013H}. Radio transient surveys are susceptible to chromatic propagation effects broadly classified as dispersion and scattering, which are induced by plasmas along the line-of-sight (LOS). Pulsars and FRBs are both affected by dispersive and scattering delays in their times of arrival, and scattering in particular can significantly reduce the detectability of these radio sources when the scattering delay ($\tau$) is comparable to or greater than the pulse width. While pulsars, being mostly Galactic sources, are predominantly affected by plasma in the Galactic interstellar medium (ISM), extragalactic FRBs can experience dispersion and scattering in plasma from their host galaxies, the IGM, other galaxies intervening their LOSs, and the Milky Way. In a given ionized medium there is a scattering ``horizon" beyond which the detectability of radio transients plummets. In this paper, we focus on the role of scattering horizons in radio transient detection, and the subsequent use of radio transients as probes of Galactic and extragalactic plasma. 

\indent The first and foremost scattering horizon relevant to both pulsars and FRBs is the Milky Way ISM. The detection of pulsars in the inner Galaxy and even near the Galactic Center continues to be of high interest \citep{2021MNRAS.507.5053E, 2021AA...650A..95T}, not only for expanding the census of both the pulsar population and of plasma in the inner Galaxy, but also for use of pulsars as probes of general relativity and the plasma environment near Sgr A$^*$ \citep{2012ApJ...747....1L, 2013Natur.501..391E, 2019MNRAS.486..360K}. The sensitivity of pulsar timing arrays (PTAs) to nanohertz frequency gravitational waves is directly related to the number of millisecond pulsars (MSPs) in the array, and detection of new MSPs with moderate to low DMs remains a high priority for PTA collaborations \citep{2013CQGra..30v4015S, 2016ApJ...819L...6T, 2021ApJ...911L..34P}.

\indent FRB surveys face similar impediments from the Galactic scattering horizon. The Galactic latitudinal dependence of the FRB population was debated early on in their discovery, due to significant discrepancies between the number of FRBs detected at low and high Galactic latitudes \citep[e.g.][]{2014ApJ...789L..26P}. \cite{2015MNRAS.451.3278M} suggested that this apparent latitudinal dependence was the result of Galactic diffractive interstellar scintillation (DISS) boosting FRB flux densities at higher Galactic latitudes. The first Canadian Hydrogen Intensity Mapping Experiment (CHIME) FRB Catalog\footnote{\url{https://www.chime-frb.ca/catalog}} \citep{2021arXiv210604352T} has provided the largest sample to date for testing the apparent Galactic latitudinal dependence of FRBs. \cite{2021arXiv210604353J} compare the latitudinal distribution of FRBs from CHIME/FRB Catalog 1 to the survey sensitivity expected from both instrumental and propagation effects, and find that the CHIME/FRB Catalog 1 is consistent with an isotropic source distribution across the sky, suggesting that any apparent latitudinal dependence in earlier FRB samples was indeed an issue of sample size and selection effects (including scattering), rather than the intrinsic FRB source distribution. \cite{2021arXiv210604353J} also argue that scattering effects play a minimal role in determining the latitudinal dependence of the CHIME survey sensitivity, but it is likely that scattering effects are more important for surveys covering lower Galactic latitudes.

\indent FRBs are also susceptible to scattering beyond the Milky Way, but the predominant origin of extragalactic FRB scattering is actively debated. While the IGM is likely far too diffuse to contribute measurable scattering, even for sources as far as redshifts $z_{\rm s}\gtrsim 1$ \citep{2013ApJ...776..125M}, the exact amount of scattering expected from intervening galaxy halos depends heavily on their density and turbulence. Models where the halo is warm and clumpy \citep[][]{2019MNRAS.483..971V} predict that scattering from intervening halos will become significant ($\tau\gtrsim 1$ ms at 1 GHz) for source redshifts $z_{\rm s}\gtrsim1$. However, FRB LOSs confirmed to intersect intervening halos indicate negligible scattering from those halos \citep{2019Sci...366..231P, 2020MNRAS.499.4716C}, and scattering from the Milky Way halo also appears to be negligible based on the scattering budgets of a few localized FRBs with highly precise scattering measurements \citep{2021ApJ...911..102O}. \edit1{An even more stringent limit on scattering in the Milky Way halo comes from FRB 202002120E, which is localized to a globular cluster in M81 \citep{2021ApJ...910L..18B,2022Natur.602..585K} and shows no evidence of scattering down to 60 ns timescales \citep[][J.M. Cordes et al. in prep.]{2021arXiv210511446N}.} Previous studies \citep{2013ApJ...776..125M, 2018MNRAS.474..318P, 2021arXiv210710858C} have also argued that the ISMs of intervening galaxies contribute negligibly to FRB scattering due to the low likelihood of LOS intersections. Using surveys of damped Ly$\alpha$ systems, \cite{2018MNRAS.474..318P} estimated the DM and scattering contributions from the HI content of galaxy ISMs, and found that both DM and scattering would be negligible even for $z_{\rm s} > 1$. However, their results were based on observations that do not directly trace ionized gas and were limited to galaxies pre-selected to have large HI column densities. 

\indent Scattering budgets for FRBs with host galaxy localizations suggest that the pulse broadening delays of these FRBs can largely be explained by scattering within their host galaxies, while their scintillation bandwidths can be explained by scattering within the Milky Way ISM \citep{2021arXiv210801172C}. These results may contradict the \cite{2021arXiv210710858C} analysis of CHIME/FRB Catalog 1, which argues that the distribution of FRB scattering times may require contributions other than the host galaxy ISM, such as FRBs' near-source environments and/or intervening galaxy halos. The recent localization of the repeating source FRB 20190520B to a dwarf galaxy contributing significant DM and scattering \citep{2021arXiv211007418N} suggests that near-source environments may be more promising candidates for resolving these apparent discrepancies between the non-localized CHIME/FRB sample of scattering times and the scattering observed from localized FRBs. Intriguingly, the scattering distribution of CHIME/FRB Catalog 1 suggests there may be a substantial number of FRBs with scattering times $>10$ ms at 600 MHz \citep{2021ApJS..257...59A}, confirming that scattering horizons are an important consideration for FRB population studies. 

\indent The goal of this paper is to assess not only the Milky Way scattering horizon for pulsars and FRBs, but also whether scattering from host galaxies and intervening galaxies, including their ISMs and halos, could cumulatively decrease FRB detections, creating an extragalactic scattering horizon. Such a scattering horizon may have important consequences for the use of FRBs as cosmological probes, as the current dearth of known FRBs from $z_{\rm s} > 1$ remains a significant hurdle to some proposed cosmological applications, such as probing the epochs of hydrogen and helium reionization \citep{2021MNRAS.502.5134B, 2021PhRvD.103j3526B}. 

\indent The paper is organized as follows: In Section~\ref{sec:theory} we summarize the formalism relating dispersion and scattering observables to the underlying density fluctuations in an ionized medium. Section~\ref{sec:density_modeling} lays out the electron density models used to predict the scattering contributions of galaxy ISMs and halos. Throughout the paper we treat a galaxy ISM as distinct from its circumgalactic medium (CGM; which we often refer to simply as the halo), due to the large dissimilarities between the structure and turbulence of these media.  Section~\ref{sec:MWscatt} describes the Milky Way scattering zone of avoidance, including its expected spatial distribution and dependence on radio pulse width. In Section~\ref{sec:nearby} we assess the scattering contributions of nearby galaxies within $\sim 100$ Mpc of the observer, by adapting the electron density modeling to galaxies in the Gravitational Wave Galaxy Catalog (GWGC). Section~\ref{sec:distant} extrapolates this analysis to distant galaxies, and Section~\ref{sec:hosts} assesses the characteristic scattering that may be expected from FRB host galaxies. The results of Sections~\ref{sec:MWscatt} through~\ref{sec:hosts} are combined in Section~\ref{sec:comp} to assess the all-sky radio scattering horizon of FRBs located at redshifts $0.5 \leq z_{\rm s} \leq 5$, including the scattering contributions of host galaxies, intervening galaxies, and the Milky Way. \edit1{Simulated DM distributions that include the IGM are also provided.} Key findings of the paper are discussed in Section~\ref{sec:discussion} and conclusions in Section~\ref{sec:conclusions}.

\section{Dispersion and Scattering}\label{sec:theory}

The chief observables in this study are the integrated electron column density ($n_e$), i.e. the dispersion measure $\DM = \int n_e dl/(1+z)$, the pulse broadening time $\tau$ (which we refer to interchangeably as the scattering time), the scintillation bandwidth $\nud$, and the angular broadening $\thetad$. These dispersive and scattering effects arise from density fluctuations in ionized gas along the LOS, which we assume follow a power-law wavenumber spectrum $\Pne(q) = \cnsq q^{-\beta} \exp(-(q\linner / 2\pi)^2)$ extending over a wavenumber range $2\pi / \louter \le q \lesssim 2\pi / \linner$, where $\louter$, $\linner$ are the outer and inner scales \citep{cfrc87}. \edit1{This form of the wavenumber spectrum explicitly invokes the inner scale using an exponential cut-off, which produces similar results to models that simply use a hard cut-off at the inner scale}. We adopt a Kolmogorov spectral index $\beta = 11/3$.

\indent For a medium with homogeneous properties, the scattering time in Euclidean space is related to the DM in the lens frame by \citep{2016arXiv160505890C, 2021ApJ...911..102O, 2021arXiv210801172C}
\begin{equation}\label{eq:taudm}
    \tau(\DM,\nu) \approx 48.03 \ {\rm ns} \ A_\tau \nu^{-4} (1+z_\ell)^{-3} \Ftilde \Gscatt \DM_\ell^2, 
\end{equation}
where $\nu$ is the observing frequency in GHz, $z_\ell$ is the redshift of the scattering medium or lens, and $A_\tau$ is a constant that converts the mean scattering delay to the $1/e$ time typically estimated from pulse shapes, as described further below. The fluctuation parameter $\Ftilde = \zeta \epsilon^2/f(\louter^2 \linner)^{1/3}$ has units of (pc$^2$ km)$^{-1/3}$ and describes the degree of turbulence in a medium composed of ionized cloudlets where $f$ is the volume filling factor, $\epsilon^2$ is the variance of density fluctuations within a cloud, and $\zeta$ represents cloud-to-cloud variations in the mean density. The dimensionless geometric leverage factor $\Gscatt$ arises from the standard Euclidean weighting $(s/d)(1 - s/d)$ in the integral over $\cnsq$ along the LOS. When the source and observer are both embedded in the scattering medium, $\Gscatt\approx1/3$, but when either the source or observer is embedded in a scattering medium that has a thickness $\delta D \ll D$, the total distance between the observer and source (such as scattering of extragalactic sources either by their host galaxies or the Milky Way), then $\Gscatt\approx1$. For scattering in an intervening galaxy or halo, $\Gscatt = 2\dsl \dlo/\dso L$, where $\dsl$, $\dlo$, and $\dso$ are the angular diameter distances between the source and lens, lens and observer, and source and observer, respectively, and $L$ is the path length through the lens. 

\indent The expression in Equation~\ref{eq:taudm} gives the mean pulse broadening time, which is not necessarily equal to the $1/e$ time that is typically measured and which assumes a Gaussian scattered image, leading to an exponential scattering tail. For non-Gaussian scattered images, the mean pulse broadening time will be larger than the $1/e$ time \edit1{\citep[e.g.][]{1999ApJ...517..299L}}. As such, we include a constant factor $A_\tau \leq 1$ in Equation~\ref{eq:taudm} that converts the mean delay to the $1/e$ delay. For the remainder of our analysis we adopt $A_\tau = 1$, but the exact value of $A_\tau$ will generally depend on properties of the scattering medium, such as the inner scale $\linner$, that are not known a priori. 

\indent For single-screen scattering, $\tau$ is directly related to $\nud$ through the uncertainty principle, $C_1 = 2\pi \tau \nud$, where $C_1 = 1$ for a homogeneous medium and $C_1 = 1.16$ for a Kolmogorov medium that is uniform along the LOS \citep{1998ApJ...507..846C}. The observed angular broadening $\thetad$ can be cast in terms of the scattering diameter $\theta_{\rm s}$ for thin-screen scattering of a source at a distance $\dso$ from the observer:
\begin{equation}
    \thetad \sim \theta_{\rm s}(\dsl/\dso) = \theta_{\rm s}\bigg(1 - \frac{\dlo}{\dso}\bigg).
\end{equation}
The mean pulse broadening delay $\tau$ can then be related to the scattering diameter as \citep{2019ARA&A..57..417C}
\begin{equation}\label{eq:tautheta}
    \tau \approx \bigg(\frac{\dsl\dlo}{\dso}\bigg) \frac{\theta_{\rm s}^2}{8{\rm ln(2)}c} (1+z_\ell)^{-3},
\end{equation}
where the factor of $(1+z_\ell)^{-3}$ accounts for the redshift scaling of both $\theta_{\rm s}$ and $\tau$.

\indent Equations~\ref{eq:taudm}-\ref{eq:tautheta} apply to the strong scintillation regime for multipath propagation, which satisfies the condition $\tau > \tau_{t} \propto \nu_t^{17/5}$, where $\nu_t$ is the transition frequency between strong and weak scattering \citep{1990ARAA..28..561R, 2002astro.ph..7156C}. At 1 GHz, the transition to weak scattering occurs when $\tau \approx 0.16$ ns. 

\indent Plasma can also reduce radio transient detection by free-free absorption. The free-free optical depth is \citep{2019ARA&A..57..417C}
\begin{equation}
    \tau_{\rm ff} = \frac{3.37\times10^{-3}}{T_4^{1.3}\nu^{2.1}}\frac{\zeta(1+\epsilon^2)}{fL_{\rm pc}}\bigg(\frac{\rm DM}{100\ {\rm pc\ cm^{-3}}}\bigg)^2,
\end{equation}
where $T_4$ is temperature in units of $10^4$ K and $L$ is the path length through the relevant plasma in pc. This effect will only be relevant in very dense environments, such as some FRB host galaxies or the inner Milky Way, and at low radio frequencies ($\nu \lesssim 300$ MHz).

\section{Electron Density Modeling}\label{sec:density_modeling}

The DM and scattering contributions of a galaxy ISM or halo depends on its electron density distribution, as summarized in Section~\ref{sec:theory}. Highly structured electron density models for the Galactic disk, including NE2001 \citep{2002astro.ph..7156C, 2003astro.ph..1598C} and YMW16 \citep{2017ApJ...835...29Y}, are calibrated using the Galactic pulsar population, and include multiple disk components and spiral arms. We use NE2001 to assess the scattering horizon of the Milky Way \edit1{disk}, \edit1{as YMW16 has been demonstrated to severely misestimate the scattering times of extragalactic LOS (see Section 3.2 of \citealt{2021ApJ...911..102O} for a detailed discussion).} A simpler electron density model is used for \edit1{the Galactic halo,} other galaxies that may intervene extragalactic-origin LOSs, \edit1{and the IGM}.
This electron density model, which separately treats the ISM, CGM, \edit1{and IGM}, is described below.

\subsection{Galaxy Interstellar Media}\label{sec:model_ISM}

The general prescription for the ISM density is of the form
\begin{equation}\label{eq:disk_model}
    n_e(z,r) = \sum_{j=1}^{k} n_{0,j} {\rm sech}^2(|z|/z_{0,j}) {\rm sech}^2(|r|/r_{0,j}),
\end{equation}
where $r$ is the galactocentric radius, $z$ is the height above the galaxy plane, $n_0$ is the mid-plane density, $r_0$ is the scale radius, $z_0$ is the scale height, and the model has $k$ components. The density model is given an integration limit defined by the ellipse $(r/r_{\rm c})^2 + (z/z_{\rm c})^2 = 1$, where $r_c$ and $z_c$ define where the density truncates, and they are kept at least four times greater than $r_0$ and $z_0$. Each model component is assigned a fluctuation parameter $\Ftilde$, which is used to evaluate the scattering time $\tau$ through Equation~\ref{eq:taudm} after integrating over $n_e(z,r)$ to obtain a DM in the galaxy frame. For the ISM of intervening galaxies we calculate $\Gscatt$ using the path length through the density model at zero impact parameter, as the change in path length through the disk is negligible compared to the distances between source, lens, and observer. 

\begin{figure}
    \centering
    \includegraphics[width=0.45\textwidth]{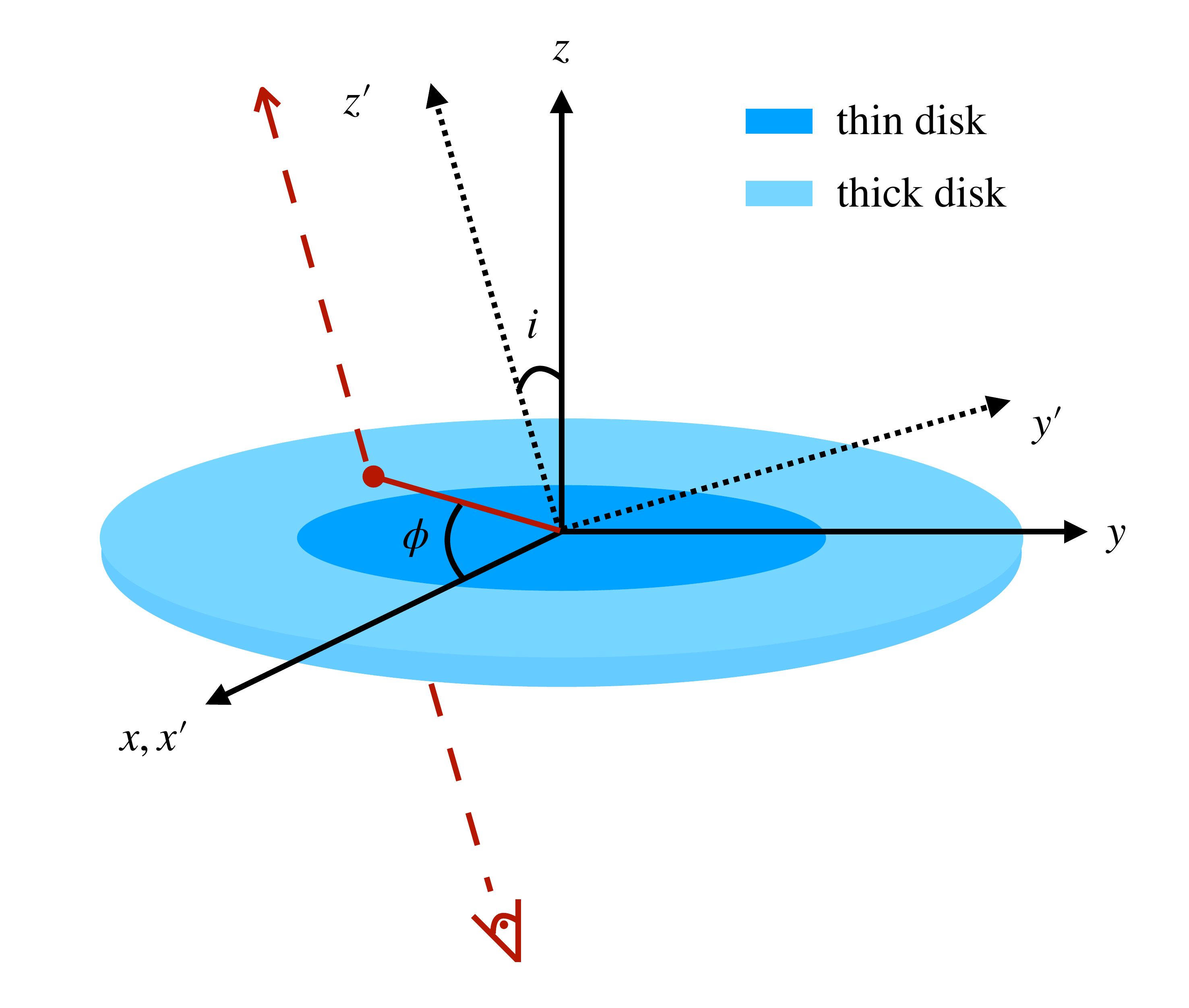}
    \caption{Schematic demonstrating the viewing geometry for a LOS intersecting a galaxy disk. The model depicted here has thin and thick disk components in galactocentric coordinates $(x,y,z)$ in the host rest frame, and coordinates $(x^\prime,y^\prime,z^\prime)$ in the observer frame. The observer frame rotates about the $x$-axis. The LOS is directed along the $z^\prime$-axis at an inclination angle $i$ relative to the galaxy frame $z$-axis and an azimuthal angle $\phi$ relative to the $x$ and $x^\prime$ axes.}
    \label{fig:disk_diagram}
\end{figure}

\indent The electron density probed by an FRB LOS through an intervening galaxy can depend heavily on viewing geometry. We characterise this viewing geometry with three parameters, the inclination angle $i$, the azimuthal angle $\phi$, and the impact parameter $r^\prime$. The density model given by Equation~\ref{eq:disk_model} is centered in the galaxy's frame, using galactocentric coordinates $(x,y,z)$ where $x^2 + y^2 = r^2$. These coordinates are transformed from the observer frame $(x^\prime,y^\prime,z^\prime)$ to the galaxy frame $(x,y,z)$ through a rotation about the $x$-axis, so that the coordinates of the galaxy frame are given by
\begin{align}
x &= x \\
y &= y^\prime{\rm cos}i + z^\prime{\rm sin}i \\
z &= -y^\prime{\rm sin}i + z^\prime{\rm cos}i.
\end{align}
The radial distance of the FRB LOS from the galaxy center is given in the observer frame by
\begin{align}
    x^\prime &= r^\prime{\rm cos}\phi \\
    y^\prime &= r^\prime{\rm sin}\phi
\end{align}
where $\phi$ is the azimuthal angle, the FRB LOS is aligned with the $z^\prime$-axis, and $r^\prime$ represents the impact parameter between the LOS and galaxy center. A diagram illustrating an example of the viewing geometry for a two-component galaxy disk is shown in Figure~\ref{fig:disk_diagram}.

\indent Below we adapt this electron density model to three characteristic galaxy ``types," spirals, dwarfs, and ellipticals (similar to approaches taken by \citealt{2015RAA....15.1629X} and \citealt{2021arXiv210710858C}). For spirals, we use two components representing a thin and thick disk, whereas for dwarfs and ellipticals we use a single, spherically symmetric density component. All models ignore galaxy cores, which despite having high densities are extremely unlikely to be intersected due to their negligible angular sizes (unless the FRB source is related to an AGN, in which case substantial scattering may be expected from the host galaxy). 

\indent Galaxy types are assigned based on the galaxy's total stellar mass: $M_*<10^8M_\odot$ are considered dwarfs, $10^8M_\odot \leq M_* \leq 10^{10.5}M_\odot$ are considered spirals, and $M_*>10^{10.5}M_\odot$ are considered equally probable of being spirals or ellipticals. These assignments are based on the galaxy stellar mass function (GSMF) observed for star-forming and quiescent galaxies out to redshifts $z\approx 3$ \citep{2021MNRAS.503.4413M}. For redshifts $z\gtrsim1.5$, constraints on the low-mass GSMF $(M_*<10^9M_\odot)$ are based on extrapolation from the observed GSMF of higher-mass galaxies. Nonetheless, the general trend for star-forming galaxies to dominate the GSMF at $M_*\lesssim10^{10.7}M_\odot$ and for quiescent and star-forming galaxies to contribute equally to the GSMF at $M_*\gtrsim10^{10.7}M_\odot$ appears to be robust despite these uncertainties. 

\indent The fiducial density model parameters for the ISM of each galaxy type are shown in Table~\ref{tab:ISM} and explained below. While these three models adopt values of $n_0$, $r_0$, $z_0$, and $\Ftilde$ designed to be suggestive of typical electron density properties for the corresponding galaxy type, both early and late-type galaxies exhibit huge diversity in terms of their ionized gas content and structure. The main purpose here is to capture the basic features of a galaxy's ISM, in order to characterize the range of scattering that may be expected for FRBs propagating through intervening galaxies. 

\begin{deluxetable*}{c | c c c c | c c c c}\label{tab:ISM}
\tabletypesize{\footnotesize}
\tablecaption{Fiducial Electron Density Parameters for a Galaxy's ISM}
\tablehead{ \colhead{} & \multicolumn{4}{c}{Component 1} &  \multicolumn{4}{c}{Component 2} \\ \cmidrule{2-5} \cmidrule{6-9}
\colhead{Galaxy Type} & \colhead{($\bar{n}_0,\sigma_{n_0}$) (cm$^{-3}$)} & \colhead{$r_0$ (kpc)} & \colhead{$z_0$ (kpc)} & \colhead{$(\Ftilde,\sigma_{\Ftilde})$ (pc$^2$ km)$^{-1/3}$} & \colhead{($\bar{n}_0$, $\sigma_{n_0})$ (cm$^{-3}$)} & \colhead{$r_0$ (kpc)} & \colhead{$z_0$ (kpc)} & \colhead{$(\Ftilde,\sigma_{\Ftilde})$ (pc$^2$ km)$^{-1/3}$} }
\startdata 
Spiral & $(0.2,0.07)$ & $5$ & $0.2$ & $(1,0.5)$ & $(0.015,0.005)$ & $10$ & $1.6$ & $(0.003,0.001)$ \\
Dwarf & $(0.05,0.017)$ & $3$ & $3$ & $(0.2,0.5)$ & \nodata  & \nodata & \nodata & \nodata \\
Elliptical & $(0.015,0.005)$ & $5$ & $5$ & $(0.003,0.001)$ & \nodata  & \nodata & \nodata & \nodata
\enddata
\tablecomments{Fiducial parameters of the electron density model for the ISM of three characteristic galaxy types: spirals, dwarfs, and ellipticals. The model parameters from left to right are the mean $\bar{n}_0$ and standard deviation $\sigma_{n_0}$ of the mid-plane density (drawn from a normal distribution), the radial scale length $r_0$, the vertical scale height $z_0$, and the mean $\Ftilde$ and standard deviation $\sigma_{\Ftilde}$ of the fluctuation parameter (drawn from a log-normal distribution). Spiral galaxies include two density components: Component 1 represents the thin disk and Component 2 the thick disk. Dwarfs and ellipticals are each modeled with a single density component. The length parameter values shown each correspond to a fiducial halo mass. For spirals and ellipticals, this fiducial mass is $1.5\times10^{12}M_\odot$ (the mass of the Milky Way), while for dwarfs the fiducial mass is $10^{9.8}M_\odot$ (similar to the mass of the SMC). Section~\ref{sec:model_ISM} describes the ISM density model in full.}
\end{deluxetable*}

\begin{figure}
    \centering
    \includegraphics[width=\linewidth]{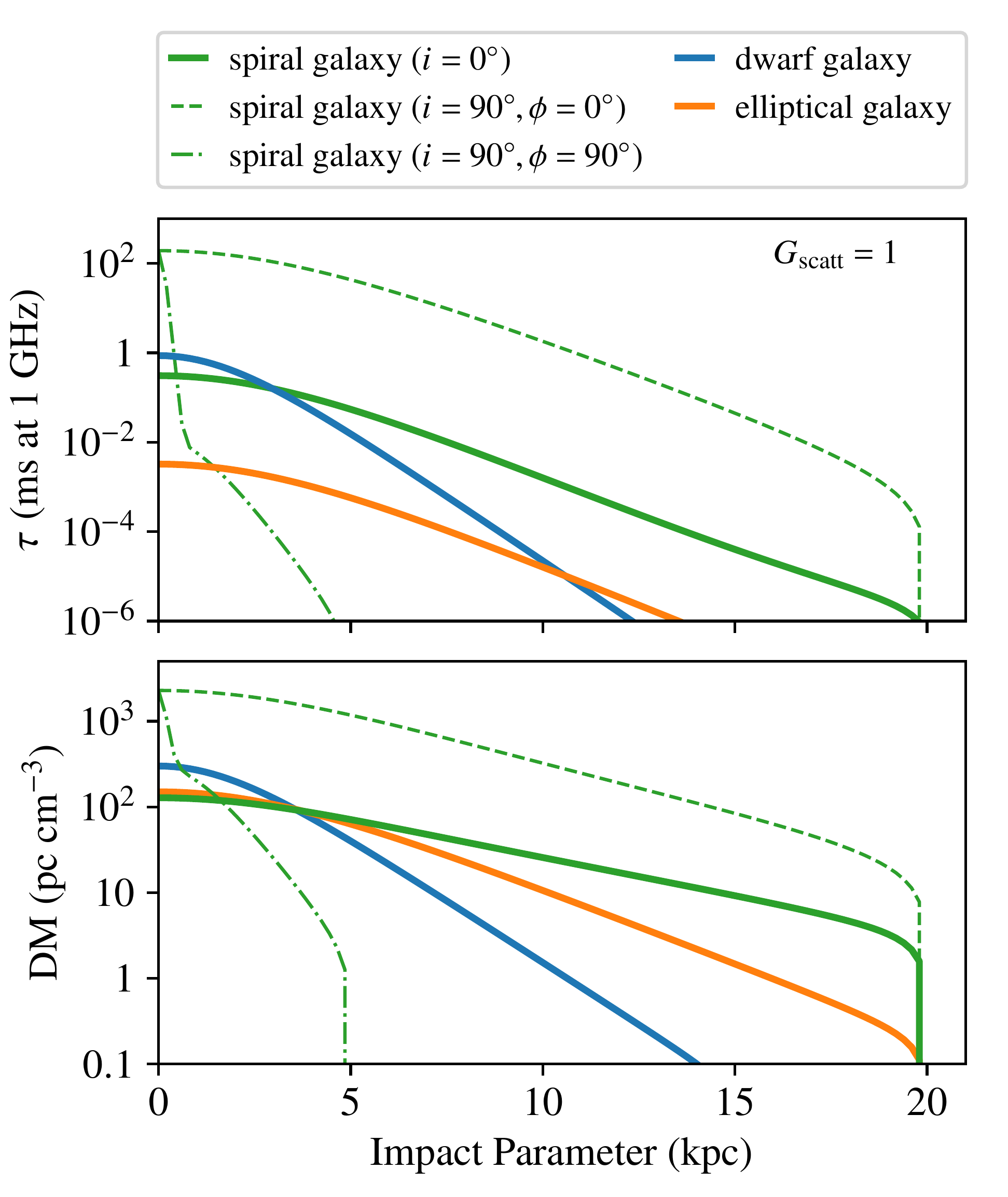}
    \caption{Scattering time in ms at 1 GHz (top) and DM (bottom) as a function of impact parameter, or radial distance from the galaxy center in kpc, for three different galaxy electron density models. Spiral galaxies are shown in green for three viewing geometries: face-on with an inclination angle $i = 0^\circ$ (thick solid line), edge-on with $i = 90^\circ$ and an azimuthal angle $\phi = 0^\circ$ (thin dashed line), and edge-on with $\phi = 90^\circ$ (thin dotted-dashed line). Dwarf galaxies are shown in blue and elliptical galaxies in orange. Each galaxy model is evaluated for its fiducial halo mass, and the mid-plane density and fluctuation parameter fixed to the mean values of their distributions (see Table~\ref{tab:ISM}). The scattering time shown is for $\Gscatt = 1$ and has not been corrected for time dilation.}
    \label{fig:galaxy_model_Gscatt1}
\end{figure}

\subsubsection{Spirals}\label{sec:spirals}

\indent The Milky Way serves as the main reference point for the electron density model of spiral galaxies in this study.  The fiducial scale height and scale radius of each component in a Milky Way-mass spiral galaxy are $(z_0,r_0)_1 = ($200 pc, $5$ kpc) for the thin disk and $(z_0,r_0)_2 = (1600$ pc, $10$ kpc) for the thick disk. These values are similar to those used in NE2001 and are based on the DM distribution of Galactic pulsars \citep{2020ApJ...897..124O}. These length parameters are scaled to spiral galaxies of different masses using the virial radius of the Milky Way halo, i.e. $r_0/r_{200} = r_{0,\rm MW}/r_{200,\rm MW}$ and $z_0/r_{200} = z_{0,\rm MW}/r_{200,\rm MW}$, where $r_{0,\rm MW}, z_{0,\rm MW}$ refer to the fiducial values for a Milky Way-mass spiral galaxy (Table~\ref{tab:ISM}), $r_{200}$ refers to the virial radius of an arbitrary galaxy halo, and $r_{200, \rm MW}$ \edit1{is the virial radius of the Milky Way. Here the virial radius is assumed to enclose the virial halo mass and is defined as the radius within which the average matter density is 200 times the cosmological critical density, $\rho_c = 3H^2/8\pi G$, where $H$ is the Hubble factor and $G$ is the gravitational constant \citep[e.g.][]{1997ApJ...490..493N}. The virial halo mass of the Milky Way is taken to be $1.5\times10^{12}M_\odot$ \citep[e.g.][]{2010MNRAS.406..264W,2019A&A...621A..56P}, yielding $r_{200, \rm MW} = 240$ kpc.}

\indent The fiducial mid-plane density for a Milky Way-mass spiral galaxy is drawn from a normal distribution with a mean $\bar{n}_0$ and standard deviation $\sigma_{n_0}$ in each disk component of $(\bar{n}_0, \sigma_{n_0})_1 = (0.2 \ \rm cm^{-3}, 0.05 \ cm^{-3})$, $(\bar{n}_0, \sigma_{n_0})_2 = (0.015 \ \rm cm^{-3}, 0.005 \ cm^{-3})$. The mean mid-plane densities are similar to those used in NE2001 and YMW16. Rather than leave $n_0$ fixed, we model $n_0$ as a distribution based on the spread of more than a factor of two that is observed in H$\alpha$ equivalent widths for galaxies of the same morphological type and mass  \citep{1983AJ.....88.1094K, 2008ApJS..178..247K}. A spread similar in magnitude is seen in the specific star formation rate for galaxies of a given mass \citep[e.g.][]{2011ApJ...730...61K}. If variations in the fluctuation parameter trace underlying changes in the star formation rate (SFR), then we might also expect $\Ftilde$ to vary between galaxies of the same mass and type. \edit1{In the Galactic thick disk alone, $\Ftilde$ varies between different pulsar LOS by $70\%$ or more \citep{2021ApJ...911..102O}.} As such, we adopt a log-normal distribution for $\Ftilde$, with a mean and standard deviation given by $(\Ftilde,\sigma_{\Ftilde})_1 = (1,0.5)$ (pc$^2$ km)$^{-1/3}$ and $(\Ftilde,\sigma_{\Ftilde})_2 = (0.003,0.001)$ (pc$^2$ km)$^{-1/3}$ for each disk component. We also assume that both $n_0$ and $\Ftilde$ distributions capture the range of values that may be encountered in spiral galaxies of different masses. A similar approach is used for both dwarfs and ellipticals, as described below. In Section~\ref{sec:comp} we explore a redshift dependence for $\Ftilde$ that follows the cosmic star formation history. \edit1{The distribution of $\Ftilde$ between different galaxies remains poorly constrained by observations, and the scattering times predicted by this density model could change substantially if the simulated fluctuation parameters do not match the true galaxy population. This effect is discussed further in Section~\ref{sec:discussion}.}

\indent While this density model is axisymmetric about the galaxy plane, different combinations of $i$, $\phi$, and $r^\prime$ yield LOS probing different sections of the disk from a variety of viewing angles. Figure~\ref{fig:galaxy_model_Gscatt1} shows the radial profiles of DM and $\tau$ for the fiducial spiral galaxy model, with $n_0$ and $\Ftilde$ fixed to the mean values of their distributions. Here $\Gscatt = 1$ and no redshift correction is applied so that the $\tau$ distribution can be simply re-scaled for different values of $\Gscatt$ via Equation~\ref{eq:taudm}. Two limiting cases are shown: $i = 0^\circ$, for which the density profile has no $\phi$-dependence, and $i=90^\circ$, for which $\phi = 0^\circ$ and $\phi = 90^\circ$ respectively correspond to LOS parallel and perpendicular to the disk mid-plane.
In Section~\ref{sec:disk_model}, DM and $\tau$ are calculated for a range of LOS through the spiral galaxy model by integrating over the density model along the $z^\prime$ axis for a range of $r^\prime$, $i$, and $\phi$.

\subsubsection{Dwarfs}

The warm ionized gas traced by H$\alpha$ emission is ubiquitous in local dwarf galaxies, although their mean H$\alpha$ luminosity is about two orders of magnitude lower than that of spiral galaxies \citep{2008ApJS..178..247K}. We base our density model for dwarf galaxies on the Magellanic Clouds, the only two other galaxies besides the Milky Way with observed pulsar populations. The observed DM distribution for the Large and Small Magellanic Clouds (LMC/SMC) spans $45$ to $273$ pc cm$^{-3}$ \citep{2013MNRAS.433..138R, 2019MNRAS.487.4332T}, exceeding the predicted DM contributions of dwarf galaxies in \cite{2015RAA....15.1629X}. Measured scattering has been reported for several pulsars in the LMC and SMC. In most cases the scattering time is inferred from the scintillation bandwidth and is between 0.1 and 1 $\mu$s at 1 GHz \citep{2022MNRAS.509.5209J}, but it is unclear whether this scattering can be attributed to the Magellanic Clouds or to the Milky Way ISM. PSR B0540$-$69 has substantial scattering that is likely associated with the pulsar's supernova remnant \citep{2003ApJ...590L..95J,2021MNRAS.505.4468G}. The scattering observed from two FRBs residing in dwarf galaxies at $z\sim0.2$, FRB 121102 and FRB 20190520B, suggests that $\Ftilde$ can range from $\lesssim0.1$ to $2$ (pc$^{2}$ km)$^{-1/3}$ in these environments (\citealt{2021arXiv210801172C, 2022arXiv220213458O}).

\indent \cite{2017ApJ...835...29Y} provide analytic density models for the LMC and SMC that were fit to their observed DM distributions. The LMC is modeled as a thick disk with a scale radius of 3 kpc and a scale height of 0.8 kpc, and contains a spherical Gaussian model of the giant HII region 30 Doradus. Fitting this model to the observed pulsar DMs (accounting for the Milky Way DM contribution), they find a mid-plane density in the LMC thick disk of $0.066\pm0.007$ cm$^{-3}$ and a mid-plane density in the HII region of $0.32\pm0.17$ cm$^{-3}$. \cite{2017ApJ...835...29Y} separately model the SMC as a spherical Gaussian with a scale radius and height of 3 kpc and fit for the mid-plane density, which they find to be $0.045\pm0.017$ cm$^{-3}$. Based on these results, the fiducial dwarf galaxy model parameters are $(\bar{n}_0,\sigma_{n_0},r_0,z_0) = (0.05 \ {\rm cm}^{-3}, 0.01 \ {\rm cm}^{-3}, 3 \ {\rm kpc}, 3 \ {\rm kpc})$ for a total halo mass $10^{9.8}M_\odot$ (similar to that of the SMC). The length parameters $(r_0,z_0)$ are re-scaled to dwarf galaxies of different masses using the virial radius of the fiducial halo mass, similar to the method used for spiral galaxies. Based on the range of $\Ftilde$ seen between the host galaxies of FRBs 121102 and 20190520B, we adopt a log-normal distribution for $\Ftilde$ with a mean and standard deviation of $0.2$ and $0.5$ (pc$^2$ km)$^{-1/3}$, respectively. This density model is used in Sections~\ref{sec:distant}--\ref{sec:comp} to assess the scattering from distant galaxies intervening FRB LOSs; however, in Section~\ref{sec:nearby} the LMC and SMC are explicitly modeled with the \cite{2017ApJ...835...29Y} prescription to assess their scattering contributions. 

\subsubsection{Ellipticals}

The ISMs of most elliptical galaxies contain hot ($T\gtrsim10^6$ K) ionized gas, although warm ($T<10^6$ K) ionized gas is also observed in a substantial fraction of these galaxies \citep[][]{1994AAS..105..341G,1996AAS..120..463M,2017ApJ...837...40P}. The spatial extent of this ionized gas varies between about a tenth to a few times the effective radius of the stellar population for different galaxies, and in most cases (at least for the most massive ellipticals) the denser warm gas tends to be concentrated near the galaxy core \citep{2017ApJ...837...40P}. Nonetheless, most ellipticals seem to have substantially less H$\alpha$ emission than late-type galaxies \citep[e.g.][]{2004AJ....127.2511N}, and are also expected to have substantially dampened turbulence compared to their spiral galaxy progenitors \citep{2021ApJ...907....2S}.

\indent Based on these characteristics, previous studies predicting the DM contribution of elliptical galaxies \citep[e.g.][]{2015RAA....15.1629X, 2021arXiv210710858C} modeled their ISM as similar to the thick disk of the Milky Way. We adopt a similar single-component density model, but again allow the mid-plane density to vary between galaxies of the same mass, giving fiducial parameters $(\bar{n}_0, \sigma_{n_0}, r_0, z_0) = (0.015 \ {\rm cm}^{-3}, 0.005 \ {\rm cm}^{-3}, 5 \ {\rm kpc}, 5 \ {\rm kpc})$ for the same fiducial halo mass as the Milky Way. While elliptical galaxies can vary dramatically in density structure, they are typically rounder than spiral galaxies leading to our simplifying assumption that $r_0 = z_0$. All length parameters are scaled for galaxies of different halo masses in the same way as for spirals, as described in Section~\ref{sec:spirals}. The $\Ftilde$ distribution is also taken to be the same as that of a spiral galaxy thick disk.

\subsection{Galaxy Halos}

\indent Models for the electron density distribution of galaxy halos typically assume that a halo's baryon content traces the underlying dark matter distribution. A variety of models have been fit to soft X-ray emission and $\rm OVI$ absorption observed from the Milky Way halo, and predict DM contributions from the Galactic halo between 10 and 120 pc cm$^{-3}$, with an average value around 50 pc cm$^{-3}$ \citep{2019MNRAS.485..648P, 2020ApJ...888..105Y, 2020MNRAS.496L.106K,2020ApJ...895L..49P}. There exist many halo density profiles based on a combination of numerical simulations and observations, but few of these are calibrated with the DMs of FRBs. We use the \cite{2019MNRAS.485..648P} (hereafter PZ19) modified Navarro-Frenk-White (mNFW) profile to estimate the DM and scattering contributions of a galaxy halo, although this model was mainly calibrated using observations of the Galactic halo, \edit1{for which it predicts a DM of 60 pc cm$^{-3}$}. We make no distinction between the halos of early or late-type galaxies beyond any difference in their halo masses.

\indent The mNFW profile gives a matter density of the form
\begin{equation}\label{eq:mnfw}
    \rho(y) = \frac{\rho_0}{y^{1-\alpha}(y_0 + y)^{2+\alpha}},
\end{equation}
where $y = K_c\times(r/r_{200})$, $r$ is radial distance from the galaxy center, and $r_{200}$ is the virial radius. The parameters $\alpha$ and $y_0$ modify the height and roll-off of the density profile. The concentration parameter $K_c$ depends on the halo mass as
\begin{equation}
    K_c = 4.67(M_{200}/10^{14}h^{-1}M_\odot),
\end{equation}
where $M_{200}$ is the virial mass and $h$ is the dimensionless Hubble constant \citep{2019MNRAS.485..648P}. The matter density profile can be converted into an electron density profile as
\begin{equation}\label{eq:neh}
    n_e(r) \approx 0.86f_{\rm b}\times(\Omega_{\rm b}/\Omega_{\rm m})\frac{\rho(r)}{m_{\rm p}}U(r),
\end{equation}
where $m_p$ is the proton mass, $\Omega_{\rm b}/\Omega_{\rm m}$ is the ratio of the baryonic matter density to the total matter density ($\Omega_{\rm b}/\Omega_{\rm m} = 0.16$ today), $f_b = 0.75$ is the adopted fraction of the galaxy's baryonic matter that is in the halo, $\rho(r)$ is given by Equation~\ref{eq:mnfw}, and we have assumed a gas of fully ionized hydrogen and helium. The function $U(r) = (1/2)\{1-{\rm tanh}[(r-2r_{200})/w]\}$ imposes a physical roll-off to the density profile at twice the virial radius over a region of width $w = 20$ kpc. While some studies choose to cut the density profile off at $r_{200}$, it is possible for ionized gas in the halo to extend beyond this boundary \citep[e.g.][]{2014ApJ...789....1D, 2020ApJ...900....9L}. For galaxy clusters, which constitute the largest halo masses in this study, virial radii defined by overdensities of $500$ times the critical density are more common, and would result in a smaller radial cutoff than the one employed here. The consequences of modifying a halo's radial extent in the analysis are noted throughout this paper when applicable.

\indent To calculate the DM and scattering contributions from a halo intervening an FRB LOS, we assume the halo is spherically symmetric and draw $\Ftilde$ from a log-normal distribution with a mean and standard deviation both set to $10^{-4}$ (pc$^2$ km)$^{-1/3}$. This distribution is based on the observed scattering of FRB 181112  \citep{2020ApJ...891L..38C} and FRB 191108 \citep{2020MNRAS.499.4716C}, both of which give upper limits on $\Ftilde$ for halos identified to intervene their LOSs \citep{2021ApJ...911..102O}. FRB 20200120E passes through the halos of both M81 and the Milky Way and shows a negligible amount of scattering \citep{2021arXiv210511446N}, \edit1{consistent with $\Ftilde \lesssim 10^{-3}$ (pc$^2$ km)$^{-1/3}$ (J.M. Cordes et al. in prep.). For the Milky Way halo, this suggests $\tau < 1$ $\mu$s at 1 GHz, and hence we exclude scattering in the Milky Way halo in the rest of the analysis. However, the DM contribution of the Milky Way halo is included to evaluate the total, predicted DMs of the FRB population.} In Section~\ref{sec:comp} we examine the effects of allowing $\Ftilde$ to vary with redshift.   

\subsection{Intergalactic Medium}

\edit1{The IGM can contribute significantly to the DM budgets of FRBs, but scattering in the IGM appears to be negligible, based both on modeling \citep[e.g.][]{2013ApJ...776..125M} and the lack of a correlation between observed scattering times and extragalactic DMs \citep{2021ApJS..257...59A, 2021arXiv210710858C}. We therefore set $\Ftilde_{\rm IGM} = 0$, but nonetheless evaluate the DM contribution of the IGM in order to simulate the total DM distribution of the FRB population. }

\indent \edit1{The mean DM contribution of the IGM is given by \citep[e.g.][]{2003ApJ...598L..79I,2004MNRAS.348..999I,2014ApJ...780L..33M}
\begin{equation}\label{eq:dmigm}
    \overline{\rm DM}_{\rm IGM}(z_{\rm s}) = n_{e,0}\int_0^{z_{\rm s}} d_H(z^\prime)(1+z^\prime)dz^\prime{}
\end{equation}
where $n_{e,0} = 2.2\times10^{-7}\times f_{\rm IGM}$ is the IGM electron density at $z = 0$, defined as a fraction $f_{\rm IGM}$ of the baryonic closure density evaluated using the Planck18 cosmology \citep{2020AA...641A...6P}. We adopt a value $f_{\rm IGM} = 0.8$ \citep[e.g.][]{2012ApJ...759...23S, 2018ApJ...867L..21Z,2021arXiv210801172C}. The factor $d_H(z)$ is given by
\begin{equation}\label{eq:dHz}
    d_{H}(z) = (c/H_0)[\Omega_\Lambda + \Omega_m(1+z)^3]^{-1/2}
\end{equation}
which depends on the speed of light $c$, the Hubble constant today $H_0$, the dark energy density today $\Omega_\Lambda$, and the matter density today $\Omega_m$. The expression for $\overline{\rm DM}_{\rm IGM}$ does not necessarily hold at redshifts $z\gtrsim3$ due to the extended onset of helium reionization between redshifts $3 \lesssim z \lesssim 4$ \citep[e.g.][]{2013MNRAS.435.3169C}. Evaluation of Equation~\ref{eq:dmigm} may therefore yield $\overline{\rm DM}_{\rm IGM}$ larger than the true value for the highest source redshift considered in this study, $z_{\rm s} = 5$.} 

\indent \edit1{Departures from the mean $\overline{\rm DM}_{\rm IGM}$, typically referred to as cosmic variance, arise from LOS intersections through foreground halos and large-scale structure \citep{2014ApJ...780L..33M,2015MNRAS.451.4277D,2019ApJ...886..135P, 2019MNRAS.485..648P}. In this study foreground halos are modeled independently, so we only include the mean $\overline{\rm DM}_{\rm IGM}$ when the total DM along an FRB LOS is evaluated. The $\overline{\rm DM}_{\rm IGM}$ estimated here is broadly consistent with values predicted by previous studies: we find $\overline{\rm DM}_{\rm IGM}(z = 1) = 870$ pc cm$^{-3}$, whereas \cite{2015MNRAS.451.4277D} find $\overline{\rm DM}_{\rm IGM}(z = 1) \approx 905$ pc cm$^{-3}$, \cite{2018ApJ...867L..21Z} finds $\overline{\rm DM}_{\rm IGM}(z = 1) \approx 855$ pc cm$^{-3}$, and \cite{2019ApJ...886..135P} find $\overline{\rm DM}_{\rm IGM}(z = 1) \approx 800$ pc cm$^{-3}$.}

\section{The Galactic Scattering Zone of Avoidance}\label{sec:MWscatt}

\begin{figure}
    \centering
    \includegraphics[width=0.45\textwidth]{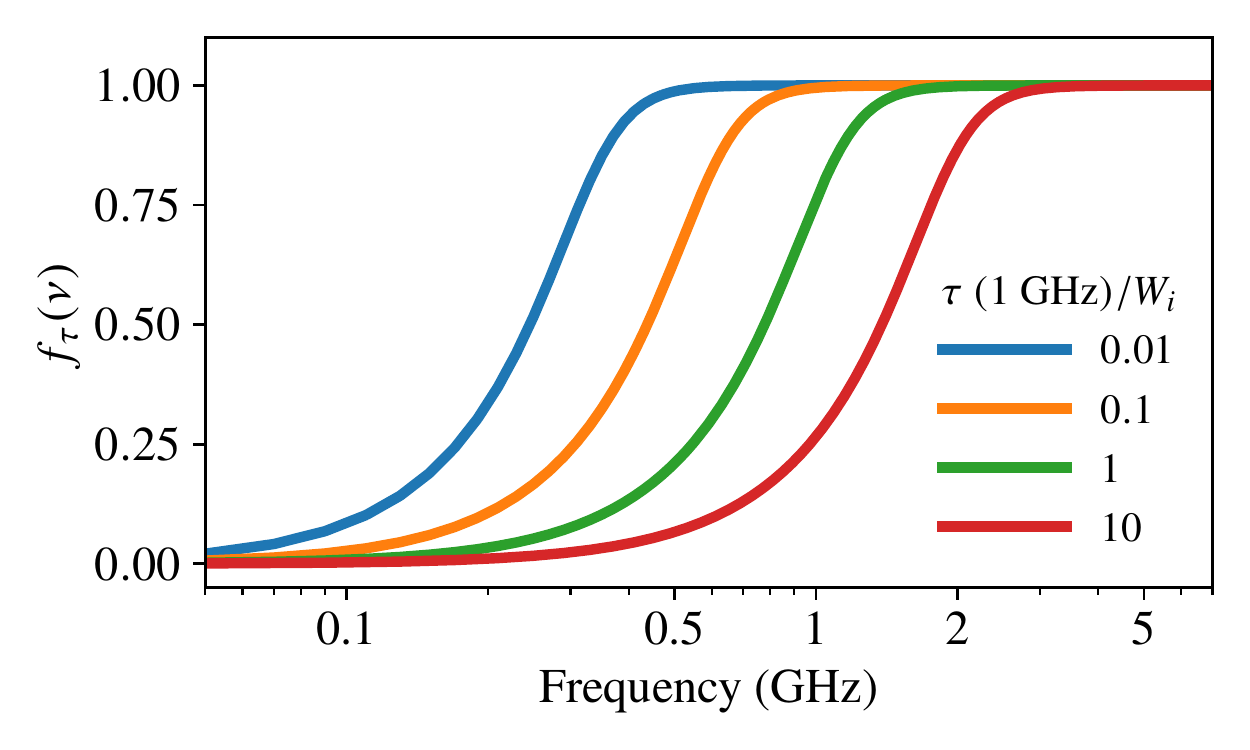}
    \caption{Reduction to the signal-to-noise ratio $f_\tau(\nu)$ due to scattering as a function of observing frequency $\nu$. The four curves correspond to different ratios of the scattering time $\tau$ to the intrinsic pulse width $W_i$ at a reference frequency of 1 GHz. Figure adapted from \cite{2019ARA&A..57..417C}.}
    \label{fig:taureduc}
\end{figure}

\indent Scattering observations of hundreds of pulsars indicate an orders of magnitude range in $\tau$ that depends heavily on Galactic latitude and longitude, in addition to observing frequency \citep[e.g.][]{2015ApJ...804...23K,2016arXiv160505890C}. Pulsar LOS towards the inner Galaxy have the largest scattering times, with PSR J1813$-$1749 exhibiting the largest currently known: $\tau \approx 0.25$ s at 2 GHz \citep{2021ApJ...917...67C}. The strong directional dependence of Galactic scattering is not only related to the electron density distribution, but also to the varying properties of turbulence in different regions of the Galaxy.

\indent We use NE2001 to quantify the Galactic scattering zone of avoidance (ZOA), the frequency-dependent region of the inner Galaxy where the predicted scattering time exceeds a given threshold. Pulse broadening will significantly reduce the signal-to-noise ratio (S/N) of observed radio transients when $\tau \geq W_i$, the intrinsic pulse width. When pulse broadening conserves fluence (which occurs when the scattering screen is sufficiently wide that rays are equally scattered towards and away from the observer; \citealt{2001ApJ...549..997C}), the resulting S/N reduction in a matched-filter output amplitude can be expressed as \citep{2019ARA&A..57..417C}
\begin{equation}\label{eq:taureduc}
    f_\tau(\nu) = [1 + 2(\tau/W_i)^2]^{-1/4}.
\end{equation}
Figure~\ref{fig:taureduc} shows the S/N reduction factor $f_\tau(\nu)$ as a function of observing frequency for different ratios of the scattering time to intrinsic pulse width $(\tau/W_i)$. The observed S/N $\rm (S/N)_{obs}$ is the product of $f_\tau(\nu)$ and the S/N of the radio spectrum. The net frequency dependence of $\rm (S/N)_{obs}$ thus depends not only on $f_\tau(\nu)$ but also on the spectral indices of the signal and the noise. In the regime where $\tau \gg W_i$, $f_\tau \propto \nu^2$ for $\tau \propto \nu^{-4}$, and for a radio source spectrum $\propto \nu^{-\alpha}$ and a noise spectrum $\propto \nu^{-\gamma}$, the net observed frequency dependence becomes $\rm (S/N)_{obs} \propto \nu^{(2-\alpha+\gamma})$. Even at $\nu \lesssim 500$ MHz where radio sources like pulsars are brighter, typical pulsar spectral indices $\alpha \approx 1.4$ \citep{2013MNRAS.431.1352B} and a noise spectrum dominated by Galactic synchrotron radiation with $\gamma \approx 2.6$ \citep{2022MNRAS.509.4923I} yields $\rm (S/N)_{obs} \propto \nu^{3.2}$. Thus even in cases where negligible scattering is detected at 1 GHz, substantial S/N reduction can still be seen at lower observing frequencies. In the following sections we consider the Galactic scattering ZOA for both Galactic-origin and extragalactic radio transients. 

\subsection{Galactic Transients}

\begin{figure*}
    \centering
    \includegraphics[width=0.8\textwidth]{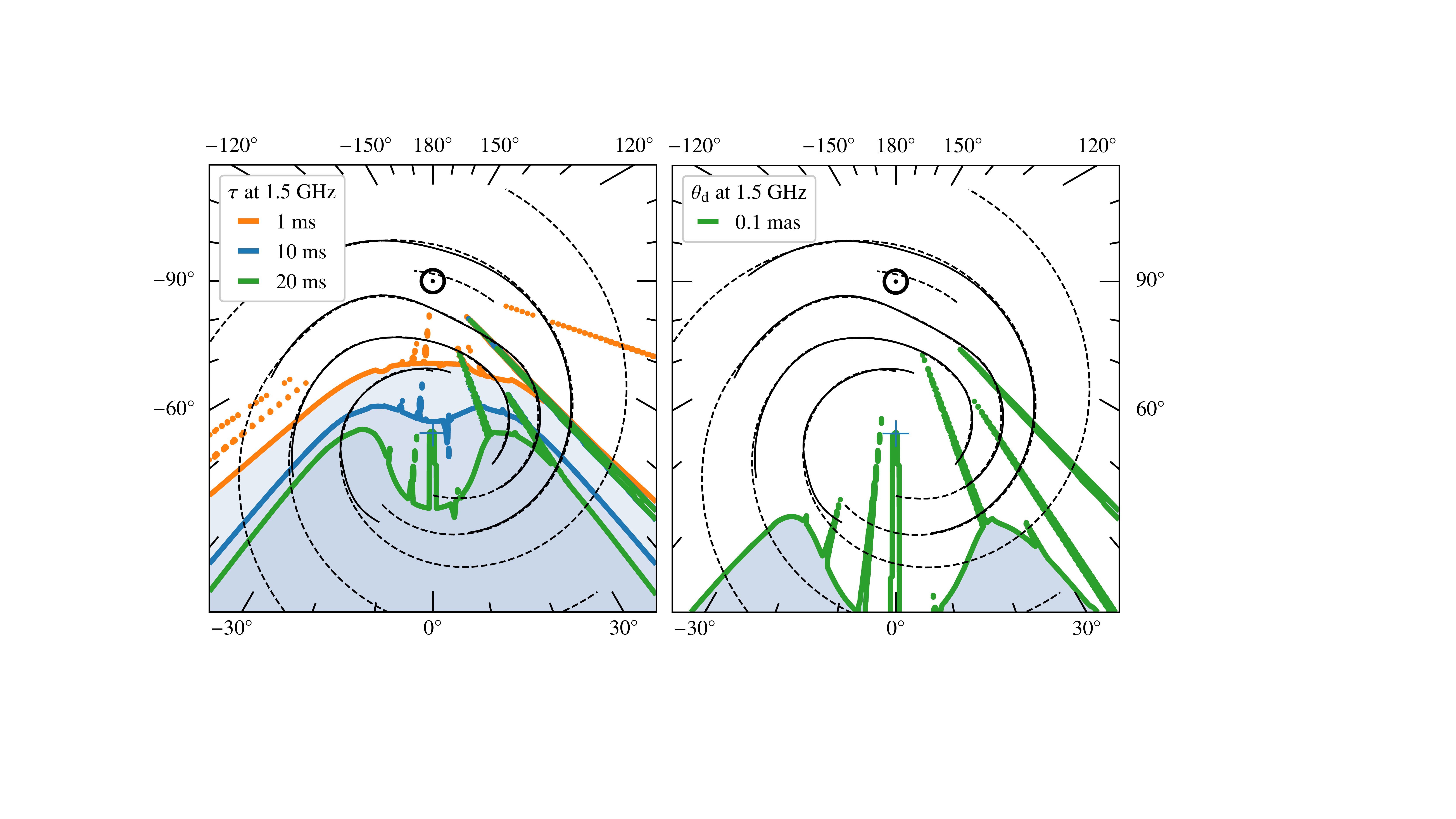}
    \caption{Scattering time $\tau$ (left) and angular broadening $\thetad$ (right) predicted by NE2001 at 1.5 GHz for Galactic pulsars, looking down on the Milky Way plane. Spiral arms are shown for both the NE2001 and \cite{1993ApJ...411..674T} electron density models by the dashed and solid black lines, respectively. The position of the observer is circled in black. Three contours are shown for $\tau$ and one contour for $\theta_{\rm d}$. NE2001 was evaluated for a pulsar population uniformly distributed across the Galactic disk with a population radius of 12 kpc and scale height of 0.5 kpc.}
    \label{fig:tau_theta_contour}
\end{figure*}

For Galactic sources like pulsars, RRATs, and magnetars, the scattering ZOA depends not only on the LOS coordinates but also on the source's distance. Figure~\ref{fig:tau_theta_contour} shows the scattering time and angular broadening distributions predicted by NE2001 for Galactic pulsars projected onto the Milky Way plane. Scattering times $\tau \geq 1$ ms at 1.5 GHz, equivalent to $\tau \gtrsim 5$ ms at 1 GHz, are reached at distances $D \gtrsim 4$ kpc from the observer for longitudes within $30^\circ$ of the Galactic Center, and $\tau \gtrsim 100$ ms at 1 GHz for $D\gtrsim 8$ kpc. At an observing frequency of 1.5 GHz, Equation~\ref{eq:taureduc} and Figure~\ref{fig:taureduc} indicate that for the majority of pulsars with intrinsic burst widths $W_i < 100$ ms, LOS through the inner Galaxy can be affected by S/N reductions as large as $50\%$, and this S/N reduction will be especially severe for time-domain surveys targeting MSPs. 

By contrast, the angular broadening distribution shown in Figure~\ref{fig:tau_theta_contour} suggests that $\thetad \gtrsim 0.1$ mas at 1.5 GHz for sources at $D \gtrsim 8$ kpc, except for a few LOS exceptions between $l \approx 20^\circ$ and $l \approx 50^\circ$ associated with discrete clumps. This $\thetad$ distribution is consistent with previous predictions that radio imaging surveys can probe further into the inner Galaxy at the same radio frequencies where scattering renders pulses undetectable in the time domain \citep[e.g.][]{1997ApJ...475..557C}, and thus far radio imaging has indeed yielded more pulsar candidates than time domain searches within a few parsecs of the Galactic Center \citep{2020ApJ...905..173Z, 2022ApJ...927L...6Z, 2022arXiv220300036S}.
However, the relatively small $\tau$ for the Galactic Center magnetar J1745$-$2900 suggests that these scattering effects may be significantly smaller than the NE2001 prediction, or that they are highly LOS-dependent \citep{2014ApJ...780L...2B, 2014ApJ...780L...3S}. 

The observed angular broadening of active galactic nuclei (AGN) places additional constraints on the Galactic ZOA. Using archival VLBI observations of about 10,000 AGN distributed across the sky between $2 \leq \nu \leq 86$ GHz, \cite{2022arXiv220104359K} find a strong latitudinal dependence of $\thetad$, with on average $\thetad > 5$ mas at 2 GHz for $|b| \leq 5^\circ$ and $|l| \leq 120^\circ$, and a median value of $\thetad \approx 2$ mas at 2 GHz for $|b| \leq 10^\circ$. To apply these $\thetad$ measurements of AGN to Galactic sources, two main effects need to be accounted for: 1) Extragalactic radio waves incident on the Milky Way are essentially planar, and hence are scattered more by Galactic plasma than spherical wave sources like pulsars \citep{2016arXiv160505890C}; and 2) the observed angular broadening is weighted by a geometric factor $\dsl/\dso$ that is about unity for extragalactic sources and $< 1$ for Galactic sources. For a single plasma screen in the Milky Way scattering both extragalactic and Galactic sources, we have
\begin{equation}
    \theta_{\rm d,gal} \approx (\theta_{\rm d,xgal}/\sqrt{3})(\dsl/\dso) < \theta_{\rm d,xgal}/\sqrt{3}
\end{equation}
where $\theta_{\rm d,gal}$ is the observed angular broadening of the Galactic sources, $\theta_{\rm d,xgal}$ is the observed angular broadening of the extragalactic sources, and $\dsl$ and $\dso$ refer to the Galactic source-screen and source-observer distances. The mean value of $\thetad$ for AGN seen through the Galactic ZOA therefore implies $\theta_{\rm d,gal} < 3$ mas for Galactic pulsars at 2 GHz. However, the observed angular broadening of J1745$-$2900 and Sgr A* indicates that $\theta_{\rm d,gal}$ can be significantly larger ($\sim300$ mas at 2 GHz; \citealt{2014ApJ...780L...2B}) near the Galactic Center. 

\subsection{Extragalactic Transients}

\begin{deluxetable}{C | C C C C | C C C C}\label{tab:scatt_zone}
\tablecaption{Extent of the Galactic Scattering Zone of Avoidance for Fast Radio Bursts}
\tabletypesize{\footnotesize}
\tablehead{\colhead{} & \multicolumn{4}{c|}{$\tau \geq 10$ ms} & \multicolumn{4}{c}{$\tau \geq 100$ ms} \\ \hline
\multicolumn{1}{c|}{$\nu$ (GHz)} & \colhead{0.4} & \colhead{0.8} & \colhead{1.4} & \multicolumn{1}{c|}{2.2} & \colhead{0.4} & \colhead{0.8} & \colhead{1.4} & \colhead{2.2}}
\startdata
$|l|_{\rm max}$  & 50^\circ & 46^\circ & 41^\circ & 37^\circ & 48^\circ & 41^\circ & 35^\circ & 27^\circ\\
$|b|_{\rm max}$   & 4.1^\circ & 2.3^\circ & 1.3^\circ & 0.5^\circ & 2.5^\circ & 1^\circ & 0.25^\circ & 0.25^\circ \\
\enddata
\tablecomments{Maximum Galactic longitude and latitude of lines-of-sight for which NE2001 predicts scattering times $\tau \geq 10$ ms and $\tau \geq 100$ ms when integrated through the entire Galaxy.}
\end{deluxetable}

\begin{figure}
    \centering
    \includegraphics[width=\linewidth]{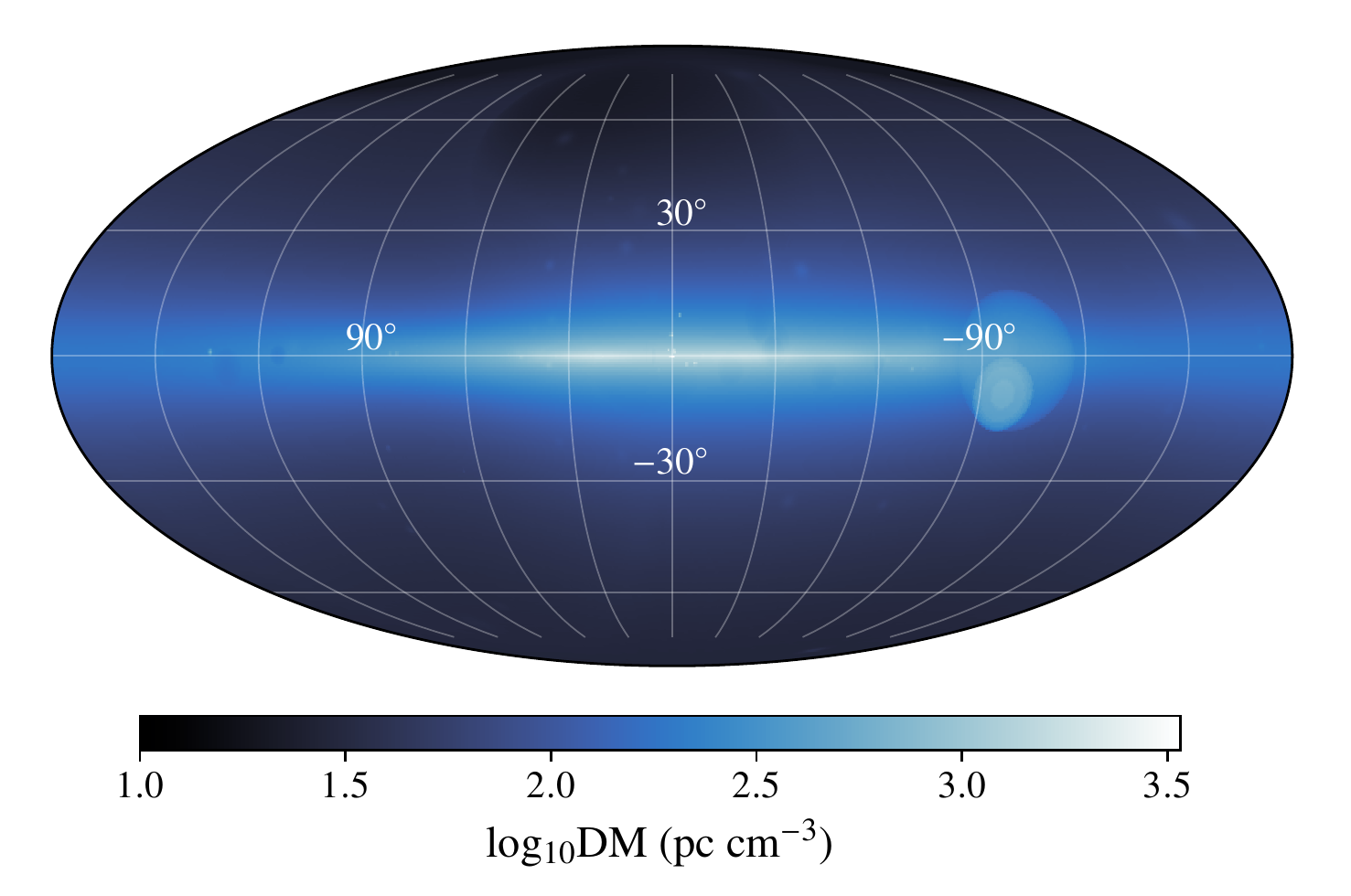}
    \caption{All-sky map (Mollweide projection) of the maximum dispersion measure (DM) from the Milky Way predicted by NE2001 \citep{2002astro.ph..7156C}, in Galactic coordinates with a spatial resolution of $0.5^\circ$.}
    \label{fig:DMne2001}
\end{figure}

\begin{figure*}
    \centering
    \includegraphics[width=0.65\textwidth]{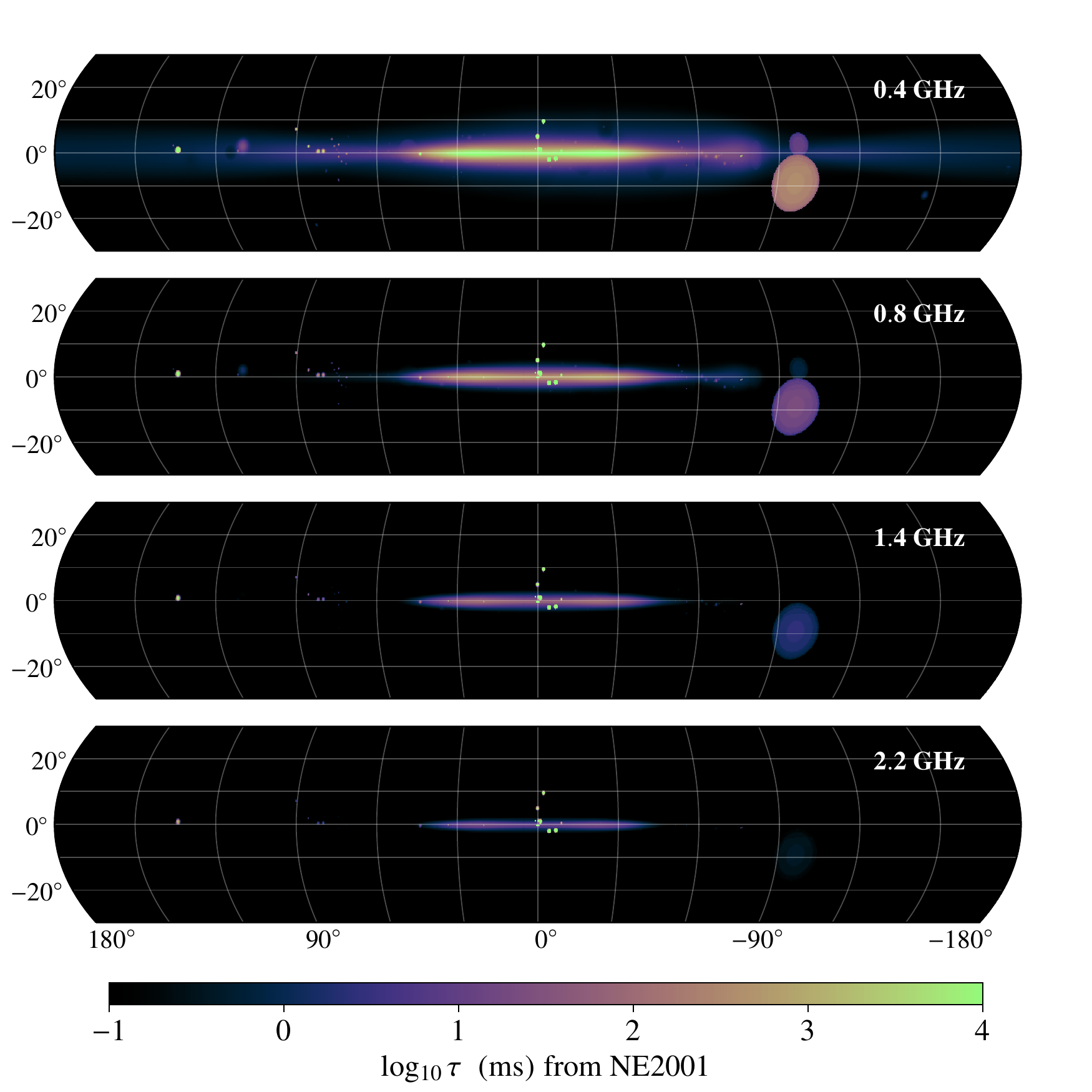}
    \caption{Mollweide projection sky map of the pulse broadening time ($\tau$) from the Milky Way predicted by NE2001 for extragalactic sources at 0.4, 0.8, 1.4, and 2.2 GHz in Galactic coordinates with a spatial resolution of $0.25^\circ$. The NE2001 scattering prediction is scaled to each reference frequency assuming $\tau \propto \nu^{-4}$, and accounts for plane wave scattering of extragalactic sources.}
    \label{fig:taune2001}
\end{figure*}

For extragalactic transients like FRBs, the Galactic ZOA depends only on LOS coordinates. Figures~\ref{fig:DMne2001} and~\ref{fig:taune2001} show maps of the DM and $\tau$ predictions from NE2001 integrated through the entire Galactic disk, with $\tau$ re-scaled to account for plane wave scattering of extragalactic sources. At 0.4 GHz, the Galactic ZOA extends over Galactic longitudes $|l| \leq 50^\circ$ and latitudes $|b|<5^\circ$. The maximum longitudes and latitudes of the ZOA are shown in Table~\ref{tab:scatt_zone} for four different observing frequencies (0.4, 0.8, 1.4, and 2.2 GHz) and two scattering thresholds, $\tau \geq 10$ ms and $\tau \geq 100$ ms. Using these four frequencies, we find that the maximum longitude of the ZOA has a roughly linear dependence on observing frequency $\nu$ of the form $|l|_{\rm max} \approx 52^{\circ}-12 \nu_{\rm GHz}$. The maximum latitude of the ZOA has a roughly quadratic frequency dependence of the form $|b|_{\rm max} \approx 0.4\nu_{\rm GHz}^{-2}$. Figure~\ref{fig:taureduc} demonstrates that for a median FRB burst width $< 5$ ms at 600 MHz \citep{2021ApJS..257...59A}, S/N reductions by more than $50\%$ may be seen at 1 GHz in the Galactic ZOA, and S/N reductions $>75\%$ at frequencies under 500 MHz. At frequencies under 100 MHz scattering delays exceed 10 ms, even at high Galactic latitudes; e.g., at 30 MHz the minimum scattering delay predicted by NE2001 for an extragalactic source is 45 ms.

\begin{figure*}
    \centering
    \includegraphics[width=\textwidth]{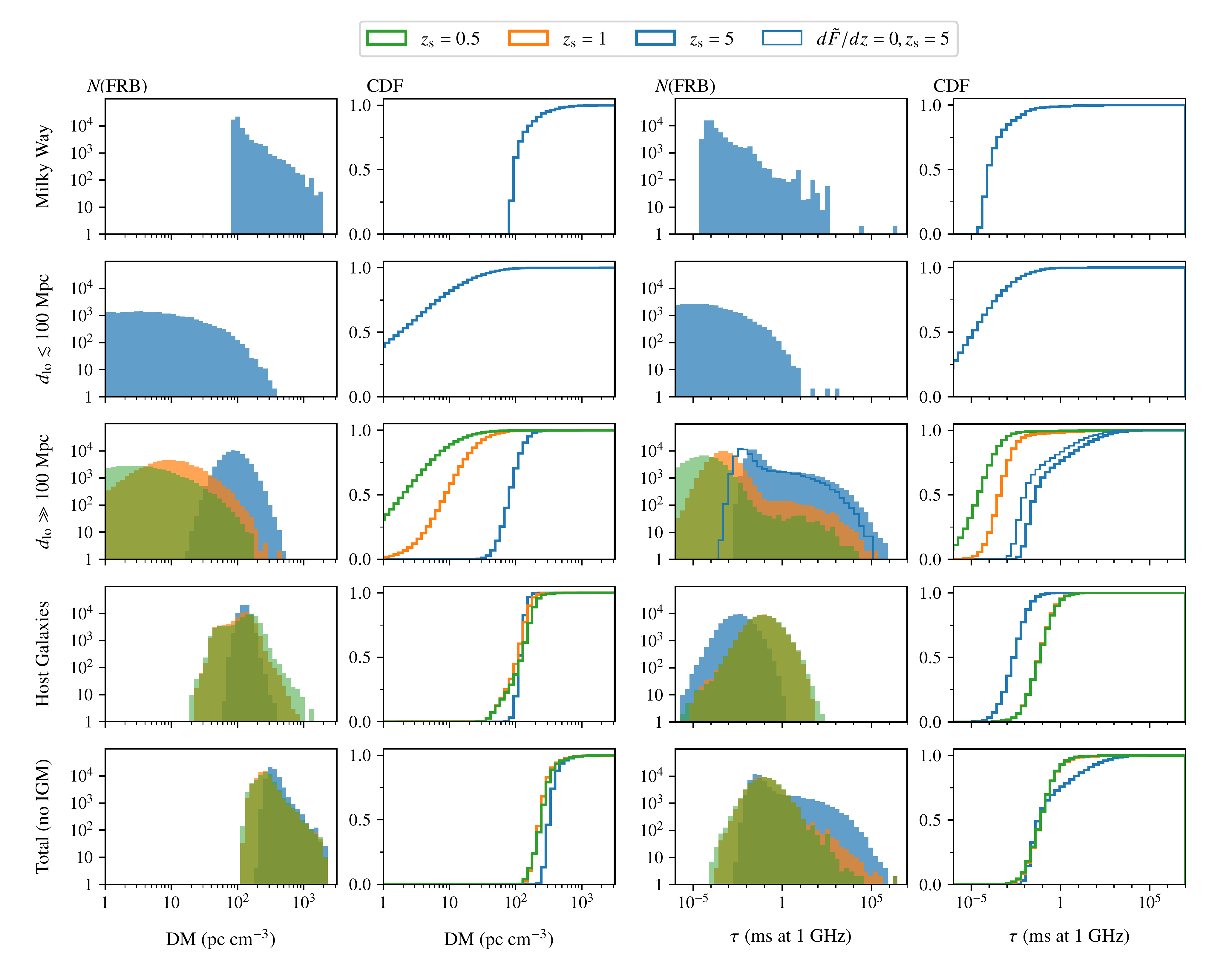}
    \caption{Histograms and cumulative distribution functions (CDFs) for \edit1{different LOS components of} DM (left) and $\tau$ (right) in the observer frame for an all-sky population of FRBs. Panels are ordered from top to bottom for each FRB LOS component considered in this study: the Milky Way, galaxies within about 100 Mpc, galaxies further than 100 Mpc, host galaxies, and the sum total of these LOS components \edit1{(excluding the IGM, which has a negligible contribution to scattering)}. Three source redshifts are shown: $z_{\rm s} = 0.5$ in green, $z_{\rm s} = 1$ in orange, and $z_{\rm s} = 5$ in blue. The Milky Way and nearby galaxy distributions do not depend on source redshift. The unfilled histogram and thin line for the $\tau$ CDF correspond to the case where $z_{\rm s} = 5$ but $\Ftilde$ is not allowed to evolve with redshift according to cosmic star formation. The host galaxy distributions shown here correspond to the case where FRBs are uniformly distributed within a single density scale height of the host ISM. \edit1{The total observable DM distribution predicted for the FRB population, including the DM contribution of the IGM, is shown in Figure~\ref{fig:dm_with_igm}.}}
    \label{fig:final_hist}
\end{figure*}

The distributions of DM and $\tau$ for $10^4$ FRBs isotropically distributed across the sky are shown in the top panel of Figure~\ref{fig:final_hist}, yielding medians and $90\%$ confidence intervals (c.i.) of $39^{+231}_{-19}$ pc cm$^{-3}$ and $1.3_{-1}^{+179} \times 10^{-4}$
ms at 1 GHz, respectively (see Section~\ref{sec:comp} for a comparison to other LOS components for FRBs). About $0.4\%$ of FRB LOS are predicted to have $\tau > 10$ ms at 1 GHz from scattering in the Milky Way alone. The Galactic scattering horizon is therefore expected to have a minimal effect on the all-sky detection rate of FRBs, but time-domain surveys targeting lower Galactic latitudes will have smaller yields, all else being equal. Source spectra that decline with frequency will partially compensate for this, but $f_\tau$ will decrease faster than the fluence increases for spectral indices smaller than 2  (for fluence $\propto \nu^{-2}$), and the increasing sky temperature below 1 GHz will exacerbate this S/N reduction.

\section{Scattering from Nearby Intervening Galaxies}\label{sec:nearby}

\begin{figure*}
    \centering
    \includegraphics[width=\textwidth]{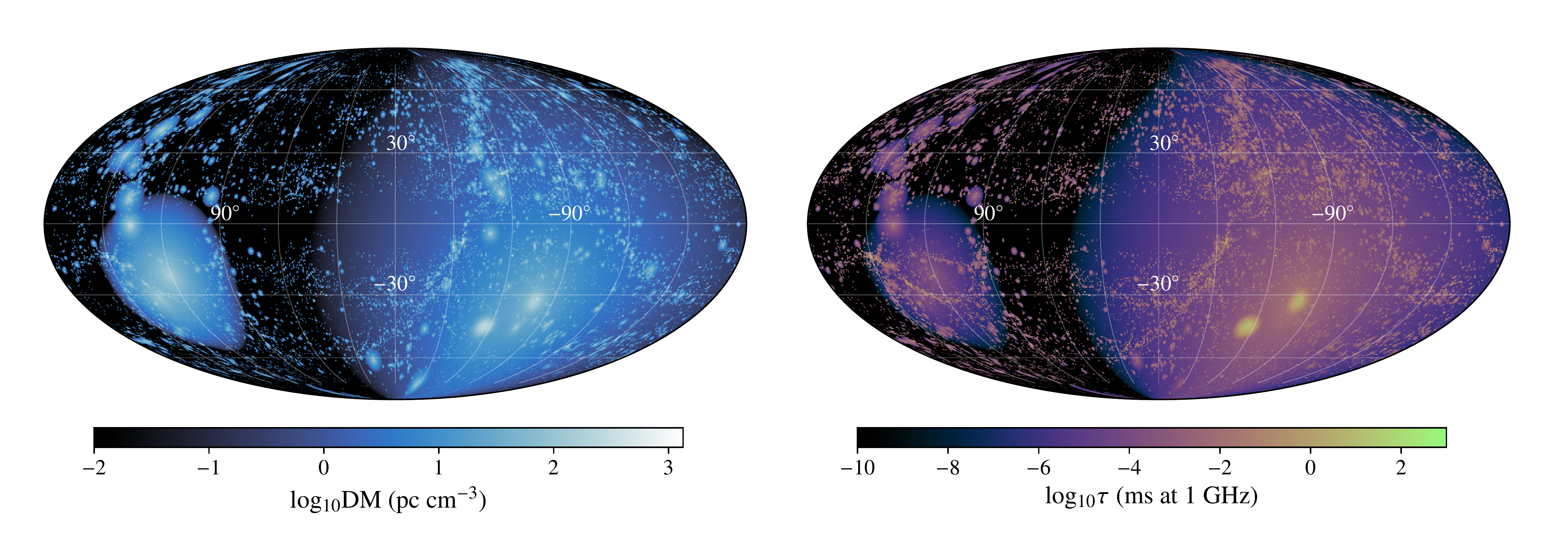}
    \caption{All-sky maps (Mollweide projection) of the predicted DM (left) and scattering time at 1 GHz (right) contributions from galaxies in the Gravitational Wave Galaxy Catalog, for FRBs at redshifts $>0.1$. The angular extents of the galaxies' halos were estimated using twice the virial radii and their distances. About 24,000 galaxies are shown with angular diameters $>0.5^\circ$. The four galaxies with the largest angular extents on the sky are M31 $(l = 121^\circ, b = -22^\circ)$, M33 $(l = 134^\circ, b = -31^\circ)$, the LMC $(l = -80^\circ, b = -33^\circ)$, and the SMC $ (l = -60^\circ, b = -44^\circ)$. The Galactic zone of avoidance is apparent as a void of galaxies near $(l=0^\circ, b=0^\circ)$.}
    \label{fig:gwgc_allsky}
\end{figure*}

\indent Low redshift ($z \ll 0.1$) galaxies can have large angular extents on the sky, making extragalactic FRB intersections with these nearby galaxies highly probable. Indeed galaxy halos at distances $<40$ Mpc appear to contribute significantly to the DMs of some FRBs in CHIME/FRB Catalog 1 \citep{2021arXiv210713692C}. Figure~\ref{fig:gwgc_allsky} shows all-sky maps of DM and $\tau$ for galaxies within $125$ Mpc, taken from the Gravitational Wave Galaxy Catalog (GWGC; \citealt{2011CQGra..28h5016W}). GWGC combines data from the Tully Nearby Galaxy Catalog, the Catalog of Neighboring Galaxies, the V8k Catalog, and HyperLEDA, and is estimated to be about $100\%$ complete out to 40 Mpc, and about $60\%$ complete out to 100 Mpc. About 24,000 galaxies are shown in Figure~\ref{fig:gwgc_allsky} and were extracted from GWGC by setting a maximum absolute $B-$band magnitude of $-18.5$ and a minimum distance of $0.5$ Mpc, thereby excluding Milky Way globular clusters and most Milky Way satellite galaxies. The Magellanic Clouds are also shown in Figure~\ref{fig:gwgc_allsky}. All galaxies are shown to their full angular extent, defined as twice the halo virial radius. 
GWGC does not provide direct measurements of galaxy masses, so we estimate the mass by scaling the galaxy's absolute $B-$band magnitude to the absolute $B-$band magnitude and mass of the Milky Way, \edit1{assuming a constant mass-to-light ratio}. For the four galaxies with the largest angular extents on the sky, the LMC, SMC, M31, and M33, we adopt independent mass measurements $M_{\rm LMC} = 1.7\times10^{10}M_\odot$, $M_{\rm SMC} = 2.4\times10^9M_\odot$ \citep{2016ARAA..54..363D}, $M_{\rm M31} = 1.5\times10^{12}M_\odot$ \citep{2012ApJ...753....8V}, $M_{\rm M33} = 10^{11.72}M_\odot$ \citep{2017AJ....154...41K}, and we note that scaling the mass by the absolute $B-$band magnitude provided by GWGC overestimates the mass of M33 by a factor of two. The DM and scattering time contributions of each galaxy were calculated using the density model described in Section~\ref{sec:density_modeling}, with each galaxy type assigned by the Hubble T-type listed in GWGC: T-types between $-6$ and $0$ were modeled as elliptical galaxies, T-types between $0$ and $9$ as spiral galaxies, and T-types between $9$ and $10$ as dwarf galaxies. The inclination angle of each galaxy was calculated directly from GWGC, while the azimuthal angle for each LOS was drawn from a uniform distribution between $0^\circ$ and $90^\circ$. The DM and scattering contributions of every galaxy along a given LOS were then summed to show the cumulative DM and $\tau$ in Figure~\ref{fig:gwgc_allsky}.

\begin{figure*}
    \centering
    \includegraphics[width=\textwidth]{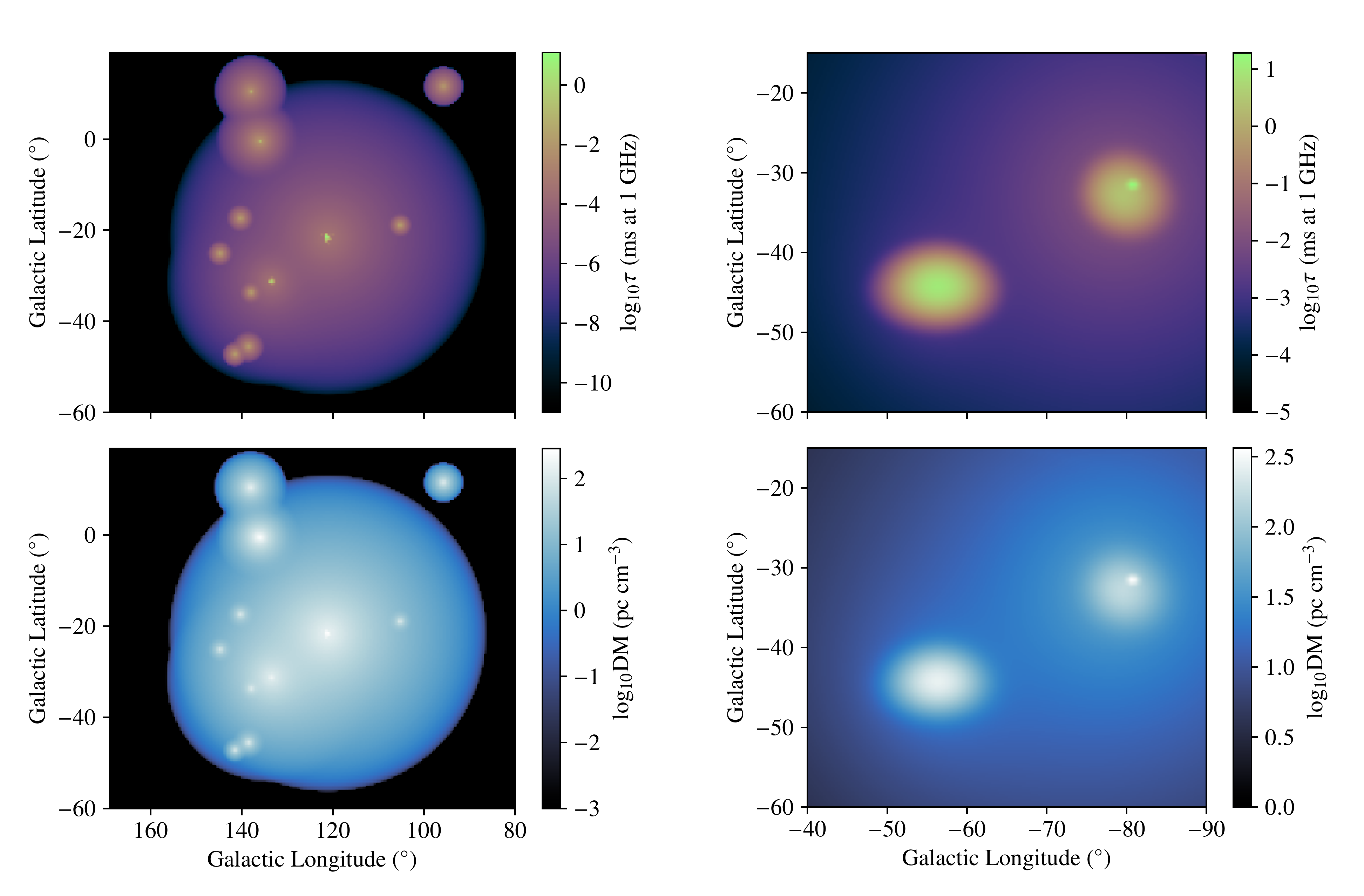}
    \caption{Scattering time $\tau$ in ms at 1 GHz (top) and DM (bottom) predicted by the density model described in Section~\ref{sec:density_modeling} for FRB LOS in the direction of M31 (left) and the Magellanic Clouds (right), and for FRB source redshifts $z_{\rm s} > 0.1$. The DM and $\tau$ shown include contributions from both the halo and disk of each galaxy. The halos of M31, M33, and their satellite galaxies overlap due to their close proximity. \edit1{The region of enhanced DM and $\tau$ near the center of the LMC is the HII region 30 Doradus.} The spatial resolution is $0.5^\circ$, whereas impact parameters $<2$ kpc from the center of M31 would be $<0.1^\circ$ on the sky. The total DM and $\tau$ can thus be even larger for impact parameters smaller than the angular resolution in this figure.} 
    \label{fig:firefly}
\end{figure*}

\indent Histograms of the total DM and $\tau$ from these nearby galaxies for an isotropic all-sky distribution of $10^4$ FRB LOS is shown in Figure~\ref{fig:final_hist}. The median and $90\%$ c.i. of the all-sky DM distribution are $2^{+35}_{-1.95}$ pc cm$^{-3}$, with about $0.6\%$ probability of DM $>100$ pc cm$^{-3}$. The scattering from these halos is extremely low (on the order of nanoseconds to microseconds at 1 GHz), although scattering in the ISMs of dwarf and spiral galaxies can be as large as 10s-100s of ms at 1 GHz depending on the impact parameter between the LOS and galaxy center. The median and $90\%$ c.i. of the all-sky $\tau$ distribution are $1.7_{-1.6}^{+2000} \times10^{-5}$ ms at 1 GHz,  with a $0.2\%$ probability of $\tau > 1$ ms at 1 GHz. Most nearby galaxy ISMs have extremely small angular extents on the sky and will not significantly reduce the all-sky rate of FRB detections. 

\indent Figure~\ref{fig:firefly} shows an expanded view of the total DM and $\tau$ in the direction of the M31 group and the Magellanic Clouds for an FRB source at $z_{\rm s}>0.1$. LOS within about $10^\circ$ of the Magellanic Clouds have $\tau \gtrsim 10$ ms at 1 GHz. This is a substantial amount of scattering that could reduce the detection of FRBs originating behind the Magellanic Clouds, and both Magellanic pulsars and FRBs detected in this direction will provide additional means for modeling the LMC and SMC. While the DM contributions of M31 and M33 are fairly large, ranging from 10s to 100s of pc cm$^{-3}$, regardless of the impact parameter between the LOS and galaxy center, the scattering contribution is much more strongly dependent on LOS location. For LOS only through the halos, $\tau \sim 1-10$s of ns at 1 GHz, whereas LOS through both the disks and halos yield $\tau \sim 0.1-10$s of ms at 1 GHz. The largest scattering time contributed by M31 through its galactic center is predicted to be $>100$ ms at 1 GHz for the same source distance, although impact parameters $<2$ kpc are not spatially resolved in Figure~\ref{fig:gwgc_allsky} or~\ref{fig:firefly}. Angular broadening measurements of AGN viewed at impact parameters $< 10$ kpc from the center of M31 show $\thetad$ at 1.6 GHz ranging from about 1 to 15 mas, the latter value corresponding to a LOS impact parameter of 0.25 kpc \citep{2013ApJ...768...12M}. Evaluating Equation~\ref{eq:tautheta} for $\thetad = 15$ mas, $\dso \gg \dlo$, and $\dlo \approx 0.8$ Mpc implies $\tau \approx 600$ ms at 1.6 MHz, affirming our model's prediction that FRB LOS at small impact parameters from M31 will be virtually undetectable, unless they are observed at $\nu \gtrsim 5$ GHz.

\indent FRB source redshift $z_{\rm s}$ has a minimal effect on these results. The distribution of $\tau$ shown in Figures~\ref{fig:gwgc_allsky} and~\ref{fig:firefly} assumes an FRB source redshift $z_{\rm s} \gtrsim 0.1$, beyond which $\Gscatt$ asymptotes to a nearly constant value that depends on the path length through an intervening galaxy and its distance. For the closest galaxies, $\Gscatt\sim1-10$ for LOS through halos because the path lengths through halos occupy significant fractions of the distances between the observer and intervening galaxies. For LOS through a galaxy ISM, $\Gscatt$ will be about two orders of magnitude larger than in the halo due to the difference in path length. 

\indent The predicted $\tau$ contributions of nearby ($z\ll0.1$) halos is consistent with recent observations that these halos appear to contribute negligibly to FRB scattering. This consistency is largely by construction because we use a nominal value of 
$\Ftilde \sim 10^{-4}$ (pc$^2$ km)$^{-1/3}$ for the CGM that is based on the small scattering times of FRBs viewed through the halos of M31, M33, and M81 \citep{2020MNRAS.499.4716C, 2021ApJ...911..102O, 2021arXiv210511446N}. While most FRBs traverse halos at large impact parameters, our fiducial spiral and dwarf galaxy models suggest that FRBs viewed at impact parameters $<2$ kpc from these galaxies' centers will be quenched by scattering. This impact parameter cutoff is fairly conservative because the electron density model does not include spiral arm structure, which can increase the amount of scattering out to impact parameters $\sim 10-20$ kpc. 

\section{Scattering from Distant Intervening Galaxies}\label{sec:distant}
For FRB sources at higher redshifts, the number of possible galaxy intersections increases, as does the geometric leverage to scattering from intervening galaxies. The corresponding scattering horizon depends not only on the location(s) of the scattering screen(s) and their electron density content(s), but also on the number of intervening galaxies. In the following sections, we estimate the probability of an FRB intersecting galaxies other than its host and the Milky Way, and we quantify a fiducial amount of scattering expected from both single intersections through a galaxy ISM or halo and from many intersections through a population of galaxies distributed along a LOS. 

\subsection{Intersection Probabilities}\label{sec:intersection_probabilities}

The mean number of galaxies encountered by an FRB from a source redshift $z_{\rm s}$ is given by \edit1{\citep[e.g.][]{2002thas.book.....P}}
\begin{equation}\label{eq:opticaldepth}
    N(z_{\rm s}) = \int_0^{z_{\rm s}} \sigma(z^\prime)n(z^\prime) \frac{d_{H}(z^\prime)}{1+z^\prime}dz^\prime,
\end{equation}
where $\sigma(z)$ is the galaxy cross-section, $n(z)$ is the number density of galaxies, and $d_H(z)$ is the Hubble expansion factor \edit1{given in Equation~\ref{eq:dHz}.}

\indent The number density of galaxies $n(z)$ can be estimated using the halo mass function (HMF), which gives the number density $n$ for a given halo mass $M$ as
\begin{equation}
    \frac{dn}{d \ {\rm ln}(M)} = f(\sigma_m)\frac{\rho_{m,0}}{M} \frac{d \ {\rm ln} (\sigma_m^{-1})}{d \ {\rm ln} (M)}
\end{equation}
where $\rho_{m,0}$ is the matter density at $z = 0$, $\sigma_m$ is the rms variance of the linear density field and depends on the linear matter power spectrum, and $f(\sigma_m)$ is a redshift-independent function of $\sigma_m$. We adopt the \cite{2008ApJ...688..709T} HMF implemented in \texttt{colossus} \citep{2018ApJS..239...35D} using the Planck18 cosmology \citep{2020AA...641A...6P}. The Tinker HMF is calibrated to redshifts $z \lesssim 2$ and accounts for redshift evolution using an overdensity threshold. 

\indent The scattering contribution of a galaxy depends heavily on its electron density distribution, with the largest scattering contributed by a galaxy ISM, and negligible scattering contributed by a galaxy halo. We therefore define separate cross-sections for halos and ISMs, using the fiducial case of a galaxy ISM confined to a disk. For halos, we adopt a circular cross-section with radius $2r_{200}$, $\sigma_{\rm halo} = 4\pi r_{200}^2$. For disks, we examine the maximum and minimum possible cross-sections $\sigma_{\rm disk}^{\rm max}$, $\sigma_{\rm disk}^{\rm min}$ corresponding to viewing a disk face-on ($i = 0^\circ$) and edge-on ($i = 90^\circ$), respectively:
\begin{align}
    \sigma_{\rm disk}^{\rm max} &= \pi r_{\rm disk}^2, \ i = 0^\circ \\
    \sigma_{\rm disk}^{\rm min} &= \pi r_{\rm disk}^2/5, \ i = 90^\circ
    \label{eq:opticaldepth_crossection}
\end{align}
where $i$ is the inclination angle and we have approximated the cross-section of an edge-on disk as an ellipse with a semi-minor axis that is $1/5$th the length of the semi-major axis $r_{\rm disk}$. Galaxy disks are assumed to trace the same number density $n(z)$ as halos, and the radius of a given disk is scaled to a given halo mass using the radii of the Milky Way disk and halo: $r_{\rm disk}/r_{200} = r_{\rm MW, disk}/r_{200, \rm MW} \approx (17 \ \rm kpc)/(236 \ kpc)$. While we assume that large halo masses ($M\sim10^{14-15} M_\odot$) contain a single disk, these masses correspond to galaxy clusters that may contain multiple disks, and the intracluster medium may be more turbulent than the CGM for a single, lower-mass halo.

\begin{figure}
    \centering
    \includegraphics[width=0.45\textwidth]{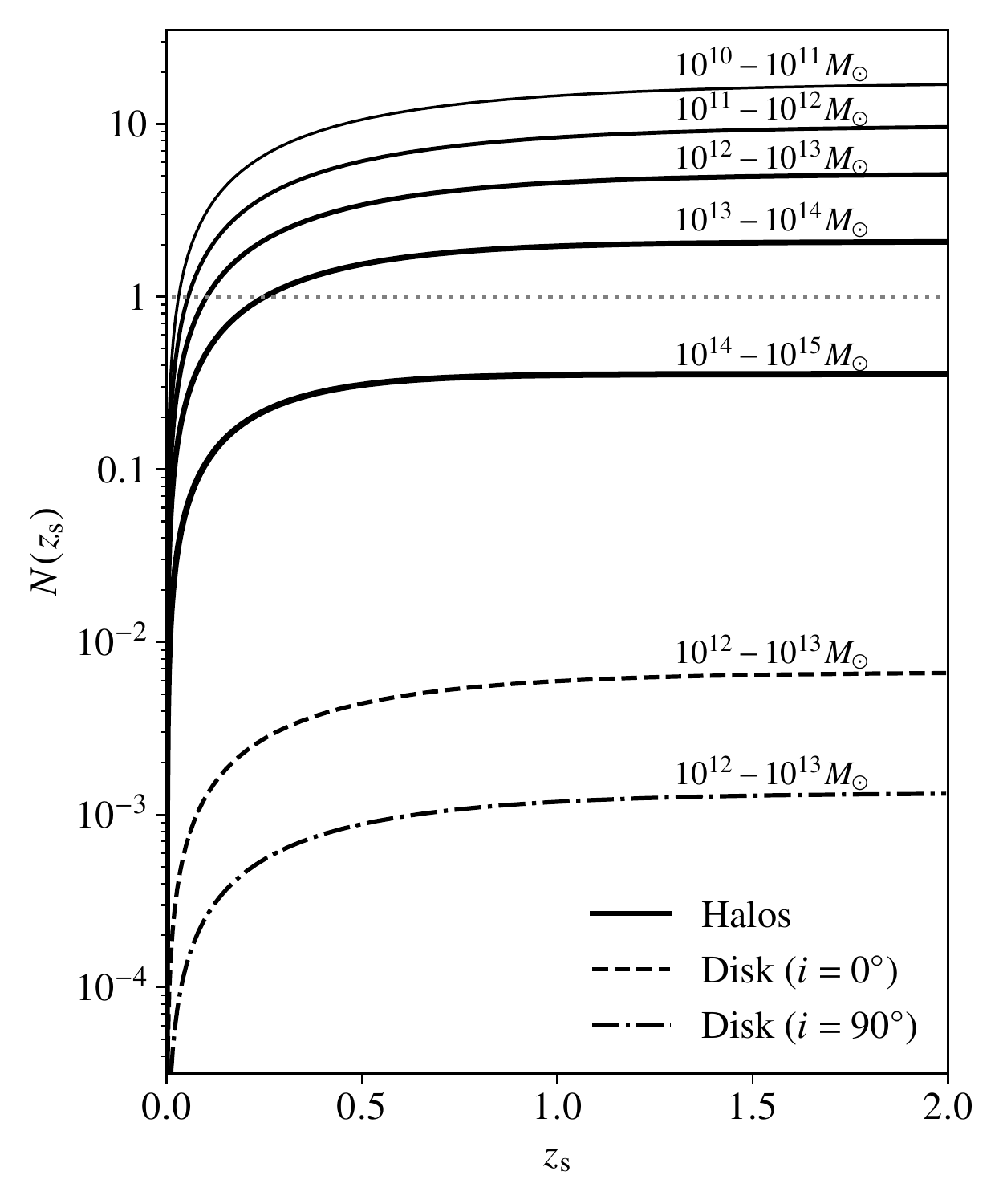}
    \caption{The mean number $N(z_{\rm s})$ of galaxy halos and disks encountered by an FRB as a function of source redshift. The intersection probability was calculated using Equations~\ref{eq:opticaldepth} through \ref{eq:opticaldepth_crossection} for halos in five different mass bins and disks in the halo mass range $10^{12}-10^{13}M_\odot$. Halos are modeled with circular cross-sections (solid lines), whereas disks are modeled for two different inclination angles: $i=0^\circ$ (dashed line) for viewing a disk face-on with a spherical cross-section, and $i=90^\circ$ (dashed-dotted line) for viewing a disk edge-on with an elliptical cross-section.}
    \label{fig:opticaldepth}
\end{figure}

\indent The intersection probability $N(z_{\rm s})$ as a function of $z_{\rm s}\leq2$ is shown in Figure~\ref{fig:opticaldepth} for five different halo mass bins between $10^{10}M_\odot$ and $10^{15} M_\odot$, spanning small halos to clusters. Intersection probabilities are also shown for galaxy disks in the $10^{12} - 10^{13} M_\odot$ mass bin for the face-on and edge-on scenarios. Previous studies \citep[e.g.,][]{2013ApJ...776..125M, 2016MNRAS.457..232C, 2019MNRAS.485..648P} typically assumed a number density $n(z)$ that is conserved with redshift and a constant cross-section, leading to predictions of substantial FRB intersections ($N\sim 0.2 - 0.4$) with clusters and large-mass halos for $z_{\rm s}\gtrsim1$. However, explicitly incorporating the redshift evolution of the HMF reveals that the number of galaxy intersections $N(z_{\rm s})$ has a steep redshift dependence for $z_{\rm s} < 0.5$, and that $N(z_{\rm s})$ for galaxy clusters asymptotes to about $0.3$ at $z_{\rm s}>0.5$. FRB intersections with lower mass halos $M < 10^{14} M_\odot$ are predicted to be $100\%$ probable for FRBs at redshifts $z_{\rm s}\gtrsim0.3$, with the lowest mass halos $M < 10^{11}M_\odot$ saturating FRB LOS at redshifts $z_{\rm s} \gtrsim 0.1$. 

\indent As expected, the probability of enountering a galaxy disk is substantially smaller due to their significantly smaller cross-sections, and the probability of encountering a face-on Milky Way-like disk asymptotes to about $5\times10^{-3}$. In reality, galaxy disks will have some distribution of inclinations, leading to values of $N(z)$ that will be distributed between the two curves shown in Figure~\ref{fig:opticaldepth} for $i = 0^\circ$ and $i = 90^\circ$. On the other hand, the predicted prevalence of low-mass halos suggests that the probability of intersecting a dwarf galaxy ISM may be as large as a few percent at $z_{\rm s} \gtrsim 0.5$.

\indent These intersection probabilities are highly sensitive to the choice of disk and halo radius. Halo intersections within one virial radius are less probable by a factor of $1/4$ than halo intersections within two virial radii. The results shown in Figure~\ref{fig:opticaldepth} can also vary depending on the choice of HMF. Variations in the assumed redshift dependence of the critical overdensity can cause the HMF to change by several percent, and deviations from universality of the HMF have been seen up to levels $\sim 10\%$ \citep[for a review, see][]{2012ARAA..50..353K}. Nonetheless, we expect that the general trend for $N(z_{\rm s})$ to asymptote at large $z_{\rm s}$ will remain the same regardless of the exact HMF chosen, due to the decrease in number density of halos at increasing redshifts. 

\subsection{Scattering from an Intervening Spiral Galaxy ISM}~\label{sec:disk_model}

\begin{figure*}
    \centering
    \includegraphics[width=\textwidth]{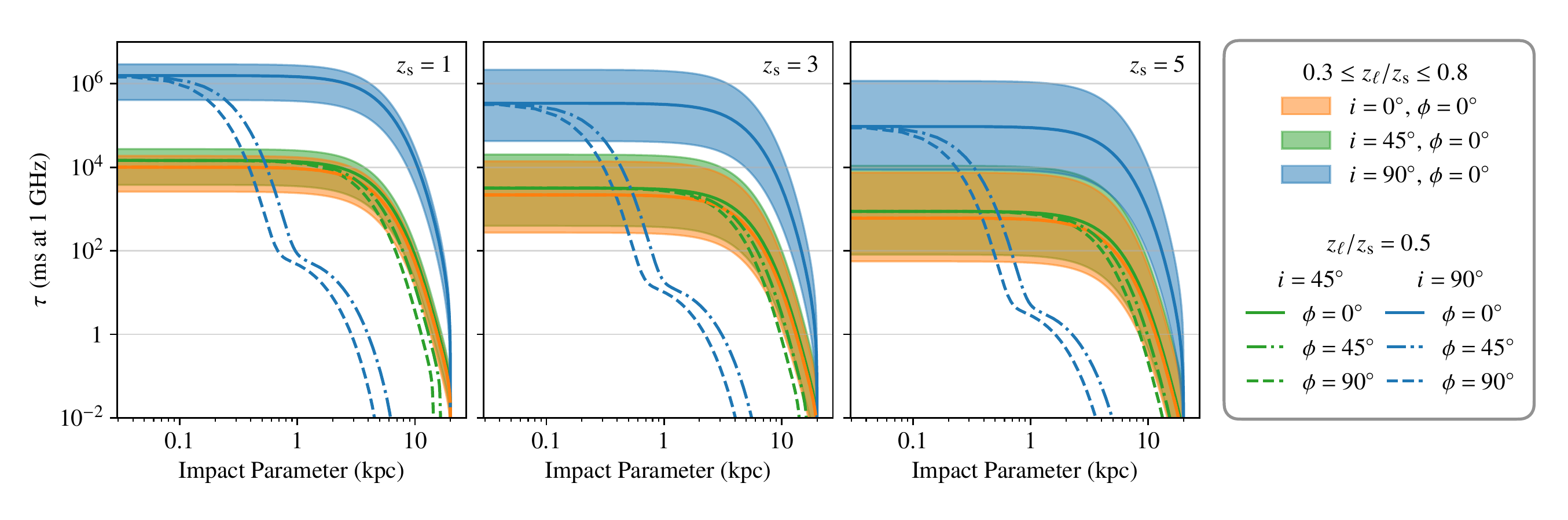}
    \caption{Scattering time in milliseconds at 1 GHz vs. impact parameter in kiloparsecs from the galaxy center, for an FRB from redshift $z_{\rm s}$ viewed through an intervening galaxy disk at a redshift $z_\ell$. The disk's electron density distribution is given by the two-component spiral galaxy model described in Section~\ref{sec:disk_model} for a Milky Way-mass galaxy, with $\Ftilde$ and $n_0$ fixed at their mean values. The orange, blue, and green curves correspond to inclination angles $i = 0^\circ, 45^\circ$, and $90^\circ$, respectively for a ratio of lens to source redshift $z_\ell/z_{\rm s} = 0.5$, while the solid, dashed, and dotted-dashed lines correspond to azimuthal angles $\phi = 0^\circ$, $45^\circ$, and $90^\circ$. The shaded regions correspond to the case $\phi=0^\circ$ for a range $0.3 \leq z_\ell/z_{\rm s} \leq 0.8$, with the inclination angles color coded as before. Results for source redshifts $z_{\rm s} = 1, 3$, and $5$ are shown from left to right.}
    \label{fig:disk_scatt}
\end{figure*}

\indent The scattering contribution of an individual spiral galaxy ISM extrapolates directly from the $\tau$ vs. impact parameter distribution shown in Figure~\ref{fig:galaxy_model_Gscatt1}, by re-scaling $\Gscatt$ from a value of $1$ to $\gg 1$ and incorporating time dilation with a factor $(1+z_\ell)^{-3}$. Figure~\ref{fig:disk_scatt} shows $\tau$ vs. impact parameter predicted by the two-component, spiral galaxy disk model for a range of $i$, $\phi$, $z_\ell$, and $z_{\rm s}$. As expected, the predicted scattering contribution of the disk has the strongest dependence on the LOS impact parameter when the disk is highly inclined, with impact parameters less than 1 kpc receiving extremely large scattering from the thin disk. For LOS viewed between $0^\circ \leq i \leq 45^\circ$ from $z_{\rm s} = 1$, the scattering contribution $\tau > 10^3$ ms at 1 GHz out to impact parameters of about 5 kpc and $\tau > 100$ ms out to impact parameters of about 10 kpc for the entire range $0.3 \leq z_\ell \leq 0.8$. The scattering contribution remains similarly large for $z_{\rm s} = 3$, but for $z_{\rm s} = 5$, the predicted scattering has a much broader distribution over the same range of $z_\ell/z_{\rm s}$. This redshift evolution mainly depends on $G_{\rm scatt}$, which tends to amplify scattering more when $z_\ell/z_{\rm s}$ is small and $z_{\rm s} > 1$ \citep{2021arXiv210801172C}. Taken alone, the results in Figure~\ref{fig:disk_scatt} indicate that even a homogeneous galaxy disk (one without spiral arms or discrete clumps) will produce so much scattering as to render most FRBs undetectable when viewed through the disk, unless the disk is viewed at an inclination angle and impact parameter large enough that the LOS only pierces the diffuse thick disk. Moreover, this result apparently applies to a broad range of source and lens redshifts. 

\begin{figure}
    \centering
\includegraphics[width=0.47\textwidth]{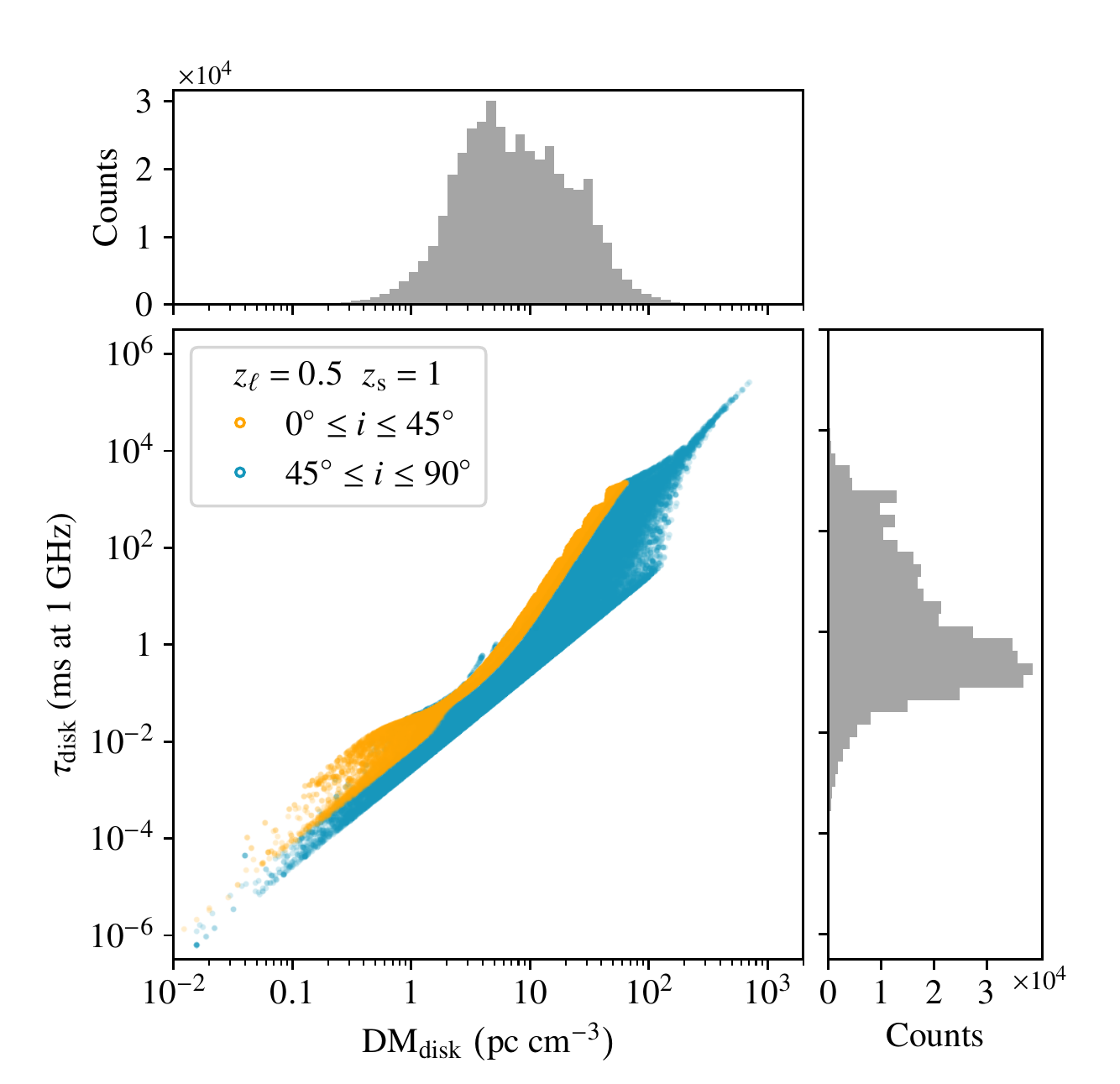}
    \caption{Scattering time in milliseconds at 1 GHz vs. DM for $10^6$ FRB LOS from source redshifts $z_{\rm s} = 1$ viewed through the spiral galaxy ISM model at a redshift $z_{\ell} = 0.5$. The LOS inclination angles are drawn from a probability density distribution $P(i) \propto {\rm sin}(i)$ for $0^\circ \leq i \leq 90^\circ$, the azimuthal angles are drawn from a uniform distribution over $0^\circ \leq \phi \leq 90^\circ$, and the impact parameters are drawn from a distribution $P(r^\prime) \propto (r^\prime)^2$ for $0 \leq r^\prime \leq 20$ kpc. Points are color coded orange for inclination angles $0^\circ \leq i \leq 45^\circ$ and blue for $45^\circ \leq i \leq 90^\circ$. The top and right-hand panels show histograms of the DM and $\tau$ produced by the intervening disk and re-scaled to the observer's reference frame.}
    \label{fig:tau_vs_dm}
\end{figure}

\indent However, whether or not this scattering actually plays a role in FRB detectability also depends on the probability of an FRB intersecting a galaxy disk with a given viewing geometry. As shown in Section~\ref{sec:intersection_probabilities}, the expected number of FRB intersections through Milky Way-mass galaxy disks is extremely small (about 1 in 1000 for $z_{\rm s} \gtrsim 0.5$). Assuming an intersection does occur, we can also estimate the expected distribution of $\tau$ and DM based on the expected distributions of $i$, $\phi$, and $r^\prime$. For $i$, we adopt a probability density distribution $P(i) \propto {\rm sin}(i)$ with $0^\circ \leq i \leq 90^\circ$; for $\phi$ we adopt a uniform distribution over the range $0^\circ \leq \phi \leq 90^\circ$; and for $r^\prime$ we adopt a distribution $P(r^\prime) \propto (r^\prime)^2$ for $0 \leq r^\prime \leq 20$ kpc, based on the cross-sectional area. Figure~\ref{fig:tau_vs_dm} shows $\tau$ vs. DM produced by the galaxy disk and re-scaled to the observer frame for $10^6$ FRB LOS generated from these distributions for $i$, $\phi$, and $r^\prime$, assuming $z_\ell = 0.5$ and $z_{\rm s} = 1$. LOS intersecting the thick disk are more likely than LOS intersecting the thin disk. The $\tau-$DM distribution in Figure~\ref{fig:tau_vs_dm} is reminiscent of the observed Milky Way pulsar $\tau-$DM relation \citep{2015ApJ...804...23K}, but Figure~\ref{fig:tau_vs_dm} shows less distinction between high and low-DM LOS because low-DM pulsars are underrepresented in the Milky Way sample and pulsar LOS towards the inner Galaxy are overrepresented in the Milky Way sample.

The median and $90\%$ confidence intervals are $\DM = (8^{+33}_{-6})/(1+z_\ell)$ pc cm$^{-3}$ and $\tau = 0.5^{+132}_{-0.46}$ ms at 1 GHz (observer frame). These values are ostensibly measurable if both the FRB host galaxy and the intervening galaxy are localized and a precise DM and scattering budget is constructed. LOS with the largest inclination angles produce DMs as large as 1000 pc cm$^{-3}$ and $\tau$ as large as $10^5$ ms at 1 GHz, in the observer frame. These intersections are extremely improbable, largely because they correspond to impact parameters less than about 1 kpc. 

\indent These results are based on a homogeneous disk model with only two components, a thin and thick disk, but spiral and bar-spiral galaxies are in reality far more inhomogeneous. Clumpy structures like spiral arms will increase the expected amount of scattering because they will contribute larger DM and can have $\Ftilde > 1$ (pc$^2$ km)$^{-1/3}$.
It is also possible that $\Ftilde$ evolves with redshift, particularly if the underlying turbulence is driven by star formation feedback or gravitational instability \citep[e.g.][]{2016MNRAS.458.1671K}. If this is the case, then the expected amount of scattering may also increase as the lens redshift increases, serving to counteract the decrease in $\Gscatt$ that occurs at sufficiently large lens redshifts. Due to the linear relationship between $\tau$ and $\Ftilde$ (see Equation~\ref{eq:taudm}), an increase in $\Ftilde$ by one order of magnitude will also increase $\tau$ by an order of magnitude (for each component of the disk).  
These caveats imply that the predicted DM and scattering contributions from the simple disk model presented here are fairly conservative. 

\subsection{Scattering from Intervening Elliptical and Dwarf Galaxy ISMs}

\indent The characteristic scattering time expected from an individual elliptical or dwarf galaxy's ISM extrapolates directly from Figure~\ref{fig:galaxy_model_Gscatt1} and the results shown for a spiral galaxy in the previous section. As shown in Figure~\ref{fig:galaxy_model_Gscatt1}, scattering in an elliptical galaxy is comparable to scattering in the thick disk of a spiral galaxy (due to the choice of model parameters), and in elliptical galaxies $\tau$ approximately tracks the the scattering of a spiral galaxy viewed face-on. The maximum $\tau$ from a dwarf galaxy lies between that of a face-on and edge-on spiral galaxy, but $\tau$ has a sharper fall-off with impact parameter for dwarf galaxies due to their smaller sizes. Scattering from elliptical and dwarf galaxies' ISMs is several orders of magnitude smaller than the maximum scattering that can be contributed by a spiral galaxy, but the relative amount that these different galaxies contribute to the cumulative scattering from a distribution of intervening galaxies for many different LOS strongly depends on the number density of these different galaxy types, as shown in Section~\ref{sec:galpop}.

\subsection{Scattering from an Intervening Galaxy Halo}

\begin{figure*}
    \centering
    \includegraphics[width=0.7\textwidth]{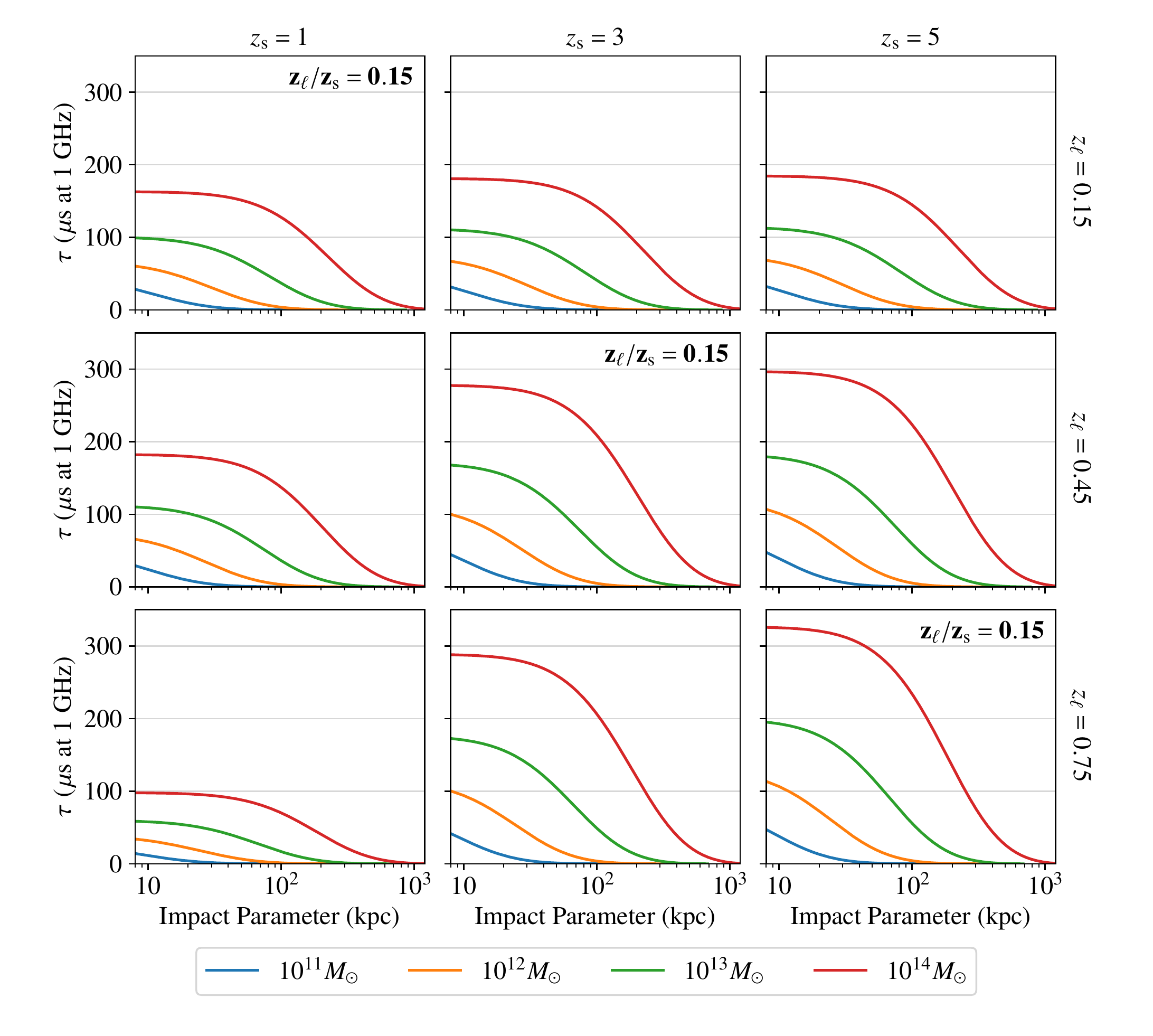}
    \caption{Scattering time in $\mu$s at 1 GHz for LOS through an individual galaxy halo at a range of impact parameters. The columns correspond to source redshifts $z_{\rm s} = 1$, 3, and 5 from left to right, and the rows correspond to an increasing halo redshift from top to bottom. The colored curves represent scattering from halos of different masses. The scattering time is calculated assuming $\Ftilde = 10^{-4}$ (pc$^2$ km)$^{-1/3}$.}
    \label{fig:haloscatt}
\end{figure*}

\indent Figure~\ref{fig:haloscatt} shows $\tau$ vs. impact parameter predicted by the mNFW profile for an individual halo intervening a LOS, for halo masses ranging from $10^{11}M_\odot$ to $10^{14}M_\odot$ and a range of $z_\ell$ and $z_{\rm s}$. Halos of mass $10^{14}M_\odot$ give the largest $\tau$ at small impact parameters, with the largest value $\tau \approx 320$ $\mu$s at 1 GHz corresponding to the case $(z_\ell,z_{
\rm s}) = (0.75, 5)$. The increase in $\tau$ with respect to halo mass is mainly due to the increase in the halo DM contribution, which outweighs any reduction in $\Gscatt$ that results from the increased path length. Large halo masses can also scatter radio emission out to much larger impact parameters, yielding $\tau \sim 10$s of $\mu$s at 1 GHz out to impact parameters between about 200 and 1000 kpc. If turbulence in the halo also evolves with redshift, which may correspond to an increase in $\Ftilde$ at larger $z_\ell$, then $\tau$ will increase linearly with respect to any increase in $\Ftilde$ (assuming $\Ftilde$ is constant across the halo). 

\indent The results shown in Figure~\ref{fig:haloscatt} will look substantially different if the halo model is only integrated out to $r_{200}$. For a smaller halo extent, $\Gscatt$ increases substantially enough with respect to DM that $\tau$ also increases. In this case, we find that $\tau$ is approximately 1.5 times larger than the values for a halo extent of $2\times r_{200}$. Nonetheless, we still find that $\tau$ is always $<500$ $\mu$s at 1 GHz from an individual halo, and that the largest $\tau$ are found at small impact parameters in the largest mass halos.

\indent FRBs at redshifts $z_{\rm s} \gtrsim 0.1$ are expected to intersect at least one low-mass halo other than that of their host and the Milky Way, and for redshifts $z_{\rm s} \gtrsim 0.3$ several halos between $10^{10}M_\odot$ and $10^{14}M_\odot$ may be intersected (see Section~\ref{sec:intersection_probabilities}). While the scattering from an individual galaxy halo may be so small as to be virtually undetectable, FRBs intersecting multiple galaxy halos may build up a cumulative amount of scattering on the order of fractions of a millisecond at 1 GHz, depending on the relative impact parameters and halo masses. This cumulative scattering may still be small compared to the amount of scattering contributed by a galaxy's ISM, but if $\Ftilde$ also evolves with redshift then multiple halo intersections may scatter FRBs enough to play a role in detection sensitivity. This possibility is explored further in the following section. 

\subsection{Scattering from a Population of Intervening Galaxies}\label{sec:galpop}

The total scattering and DM from a population of distant intervening galaxies,  evaluated for $10^4$ independent FRB LOS at source redshifts $z_{\rm s} = 0.5, 1,$ and $5$, are shown in Figure~\ref{fig:final_hist}. The mean number of galaxy intersections was calculated using the HMF for masses between $10^9M_\odot$ and $10^{15}M_\odot$ and redshifts $z < z_{\rm s}$, excluding the closest galaxies at $z \ll 0.1$ (which were already evaluated using GWGC). The intervening galaxy redshifts were drawn from a Poisson process, and the DM and $\tau$ contributed by each galaxy were calculated using the formalism laid out in Sections~\ref{sec:theory} and~\ref{sec:density_modeling}. Galaxy types were assigned by stellar mass, which was calculated using the stellar-to-halo mass relation (SHMR). We adopt the analytic approximation of the SHMR provided by \cite{2020AA...634A.135G}, who fit a redshift-dependent model of the SHMR based on both the observed GSMF from COSMOS and the $\Lambda$CDM \texttt{dustgrain}-\textit{pathfinder} simulation.

In addition, $\Ftilde$ was allowed to evolve with redshift according to the cosmic star formation history \citep[CSFR;][]{2014ARAA..52..415M},
\begin{equation}\label{eq:Ftilde_z}
    \Ftilde(z) \approx \Ftilde_0 \times \frac{(1+z)^{2.7}}{1 + [(1+z)/2.9]^{5.6}},
\end{equation}
where $\Ftilde_0$ was drawn from the fiducial PDF given in Section~\ref{sec:density_modeling}. The individual DM and $\tau$ contributions of each intervening galaxy were re-scaled to the observer frame before summing all of the intervening galaxies' contributions, yielding the total DM and $\tau$ distributions.

\indent The median and $90\%$ c.i. of DM and $\tau$ for intervening galaxies are shown in Table~\ref{tab:final_tab}. Both the DM and scattering contributions of these intervening galaxies are negligible at 1 GHz for $z_{\rm s} = 0.5$, and remain small for $z_{\rm s} = 1$. However, the predicted range of DM and $\tau$ increases dramatically for $z_{\rm s} = 5$. In this case, the median and $90\%$ c.i. for DM are $87_{-41}^{+86}$ pc cm$^{-3}$ and for $\tau$ they are $0.05_{-0.042}^{+285}$ ms at 1 GHz (observer frame). There is a $16\%$ predicted probability of $\tau > 10$ ms at 1 GHz from intervening galaxies alone. Keeping $\Ftilde$ constant with redshift reduces the expected scattering by about one order of magnitude. High-redshift FRBs get the largest scattering due to the combination of increased $\Gscatt$, more galaxy intersections, and an additional increase in $\Ftilde$ if it is redshift-dependent. The $\tau$ distributions are bi-modal regardless of source redshift, with the peaks predominantly arising from halos and the tail predominantly arising from ISMs. The scattering distributions are heavily dominated by dwarf galaxies, which comprise the largest fraction of galaxy types intersected.

\begin{deluxetable*}{l | C C | C C | C C }\label{tab:final_tab}
\tablehead{\multicolumn{7}{c}{Dispersion Measure (pc cm$^{-3}$ in the observer frame)} \\ \hline
\colhead{} & \multicolumn{2}{c}{$z_{\rm s} = 0.5$} &
\multicolumn{2}{c}{$z_{\rm s} = 1$} & \multicolumn{2}{c}{$z_{\rm s} = 5$} \\ \hline
\colhead{} & \colhead{Median} & \colhead{$90\%$ Confidence} & \colhead{Median} & \colhead{$90\%$ Confidence} & \colhead{Median} & \colhead{$90\%$ Confidence}}
\tablecaption{Dispersion and Scattering for an Isotropic All-Sky Distribution of FRBs}
\startdata
Milky Way$^\dagger$ & 102 & [83,333] & 102 &  [83,333] & 102 &  [83,333]  \\
$\dlo \lesssim 100$ Mpc & 1.9 & [0.05,37] & 1.9 & [0.05,37] & 1.9 & [0.05,37] \\
$\dlo \gg 100$ Mpc & 2 & [0.16,20] & 9 & [2.0,42] & 87 & [46,173] \\
Host Galaxies$^{\dagger \dagger}$ & 132 & [45,237] & 116 & [46,196] & 127 & [96,170]  \\
Total (includes IGM)$^{\dagger\dagger\dagger}$ & 685 & [577,935] & 1118 & [1027,1353] & 4010 & [3935,4247] \\ \hline
\multicolumn{7}{c}{\rm Scattering Time (ms at 1 GHz in the observer frame)} \\
\hline
Milky Way & 1.3\times10^{-4} & [4\times10^{-5},0.018] & 1.3\times10^{-4} & [4\times10^{-5},0.018] & 1.3\times10^{-4} & [4\times10^{-5},0.018] \\
$\dlo \lesssim 100$ Mpc & 1.7\times10^{-5} & [1.8\times10^{-8},0.019] & 1.7\times10^{-5} & [1.8\times10^{-8},0.019] & 1.7\times10^{-5} & [1.8\times10^{-8},0.019] \\
$\dlo \gg 100$ Mpc & 3.2\times10^{-5} & [5.6\times10^{-7},1.5\times10^{-3}] & 5\times10^{-4} & [3.7\times10^{-5},0.014] & 0.05 & [8.3\times10^{-3},285] \\
Host Galaxies & 0.09 & [5\times10^{-3},1.8] & 0.08 & [5\times10^{-3},1.6] & 3\times10^{-3} & [2\times10^{-4},0.05]  \\
Total & 0.1 & [6\times10^{-3},2.4] & 0.1 & [7\times10^{-3},2.7] & 0.08 & [0.01,298] \\
\enddata
\tablecomments{Median values and $90\%$ confidence intervals of DM and $\tau$ (observer frame) from different LOS components for an isotropic all-sky distribution of FRBs, tested for three source redshifts $z_{\rm s} = 0.5, 1$, and $5$. $\dagger$ Based on NE2001 and including the DM contribution of the Galactic halo, which is evaluated using the halo density model described in Section~\ref{sec:density_modeling}. Scattering in the Galactic halo is considered negligible. $\dagger \dagger$ For FRBs distributed within one density scale height. $\dagger \dagger \dagger$ Total DM includes the mean DM contribution of the IGM at each source redshift, evaluated using Equation~\ref{eq:dmigm}. Scattering in the IGM is considered negligible.}
\end{deluxetable*}

\section{Scattering in Host Galaxies}\label{sec:hosts}

\indent Most FRBs propagate through some portion of their host galaxies, although how much dispersion and scattering they experience in their host galaxies will depend heavily on their locations within the hosts and the galaxy structure. While about half of the FRBs with published localizations show evidence of significant scattering from their host galaxies \citep{2021arXiv210801172C}, it remains unclear whether this is a common trend in the broader observed FRB population \citep{2021arXiv210710858C}. The geometric leverage to scattering $\Gscatt$ is generally much smaller for FRBs embedded in the ionized ISM of their host galaxies than it is for scattering from an intervening galaxy located far from the host or observer. Below we extrapolate the scattering from an intervening galaxy ISM or halo to scattering in a host galaxy, and evaluate the scattering from a distribution of host galaxies at a range of redshifts.

\subsection{Intervening Galaxies vs. Host Galaxies}

\indent The chief difference between scattering in an intervening galaxy vs. scattering in a host galaxy arises from the geometric configuration of the source, lens, and observer, which impacts not only $\Gscatt$ but also the redshift corrections to DM and $\tau$. While the DM in the observer frame will re-scale with lens redshift according to the usual $1/(1+z_\ell)$ relation, $\tau$ will re-scale with redshift according to $(1+z_\ell)^{-3}$, in addition to its linear dependence on $\Gscatt$. In Section~\ref{sec:disk_model} and Figure~\ref{fig:disk_scatt} we showed the characteristic amount of scattering expected from a galaxy disk for the case $(z_\ell,z_{\rm s}) = (0.5,1)$, which corresponds to $\Gscatt \approx 10^5$. For scattering of an FRB embedded in its host galaxy ISM, $\Gscatt \approx 1$ and the DM in the galaxy rest frame is approximately half the value predicted for an intervening galaxy, if the FRB source is located halfway through the host. In this case, the ratio of scattering from the host galaxy to scattering from an identical intervening galaxy is
\begin{equation}
    \frac{\tau_{\rm h}}{\tau_{\rm int}} \approx \frac{1}{4}\bigg(\frac{1+z_{\rm h}}{1+z_{\rm int}}\bigg)^{-3} \frac{\Gscatt^{\rm h}}{\Gscatt^{\rm int}}
\end{equation}
where the sub- and superscript $\rm h$ refers to the host galaxy and $\rm int$ refers to the intervening galaxy. For reference, if we take the distribution of $\tau$ from Figure~\ref{fig:tau_vs_dm} for an intervening galaxy at a redshift $z_\ell = 0.5$, and instead consider scattering from a host galaxy with an identical density profile at a redshift $z_{\rm s} = 0.5$, then the distribution of $\tau$ shown in Figure~\ref{fig:tau_vs_dm} will be reduced by a factor of about $10^{-5}$ for $\Gscatt^{\rm h} = 1$. However, it is possible for $\Gscatt^{\rm h}$ to be greater than 1 if the FRB is offset from the scattering layer; this scenario could include FRBs that have migrated away from their galaxy disks, and/or FRBs subject to thin-screen scattering from discrete plasma structures. 

\subsection{Scattering from a Population of Host Galaxies}

The population of FRB host galaxies is poorly constrained by current observations, with only $\sim1\%$ of published FRBs associated to their hosts. Most localized FRBs currently appear to reside in spiral galaxies \citep[e.g.][]{2020ApJ...903..152H}, two lie in dwarf galaxies \citep{2017Natur.541...58C, 2021arXiv211007418N}, and one lies in an elliptical or lenticular galaxy \citep{2019Sci...365..565B}. The relationship between the current sample of known host galaxies and FRB progenitor channels also remains unclear \citep{2020ApJ...903..152H, 2021ApJ...917...75M, 2022AJ....163...69B}, although most FRB hosts currently appear to be moderately star-forming \citep{2022AJ....163...69B}. As our primary goal is to assess a fiducial amount of scattering that may be expected from host galaxies, we do not examine a variety of progenitor channels or large-scale redshift evolution in the FRB progenitor population. Instead, we make the simplified assumption that the distribution of FRB host galaxies traces the GSMF (where we again adopt the model provided by \citealt{2021MNRAS.503.4413M}). However, we note that 
the host galaxies of localized FRBs do not currently appear to trace stellar mass \citep{2022AJ....163...69B}. If FRB host galaxies do trace the GSMF, then a substantial fraction of FRBs should reside in dwarf galaxies, which does not appear to be the case. The difference between the density models used for dwarf and spiral galaxies here is small enough that modifying this assumption has a negligible impact on our results, as long as elliptical galaxies are assumed to be extremely rare FRB hosts. In both cases, $\Ftilde$ is scaled with redshift according to Equation~\ref{eq:Ftilde_z}.

As before, we simulate $10^4$ FRBs at each of three source redshifts, $z_{\rm s} = 0.5, 1,$ and $5$, and draw their host galaxy stellar masses and types using the GSMF. The galaxy halo masses are determined from the stellar mass using the SHMR provided by \cite{2020AA...634A.135G}. We consider two cases for FRB locations within their host galaxies: 1) As a fiducial scenario, we assume that the FRBs are isotropically distributed within $200\ {\rm pc} \times (r_{200}/r_{200,0})$ from their galaxy centers, where $r_{200,0}$ refers to the fiducial virial radius of the corresponding galaxy type, at the galaxy redshift. This location cutoff is equivalent to the requirement that an FRB lie within the thin disk of a spiral galaxy, and serves as a proxy for the hypothesis that FRBs are younger sources distributed near active star-forming regions. 2) We also discuss a scenario in which the FRBs are uniformly distributed throughout the galaxy ISM, which encompasses a broader range of potential progenitors, including older sources that have migrated away from galaxy disks. 

The results shown in Figure~\ref{fig:final_hist} and Table~\ref{tab:final_tab} correspond to the first scenario, in which the FRB locations are restricted to within one density scale height. The median and $90\%$ c.i. for DM (observer frame) are $132^{+105}_{-87}$ pc cm$^{-3}$ for $z_{\rm s} = 0.5$, $116^{+80}_{-70}$ pc cm$^{-3}$ for $z_{\rm s} = 1$, and $127^{+43}_{-31}$ pc cm$^{-3}$ for $z_{\rm s} = 5$. 
The median and $90\%$ c.i. for $\tau$ (observer frame) are $0.09_{-0.085}^{+1.8}$ ms at 1 GHz for $z_{\rm s} = 0.5$, $0.08_{-0.075}^{+1.6}$ ms at 1 GHz for $z_{\rm s} = 1$, and $(3_{-2.8}^{+47})\times10^{-3}$ ms at 1 GHz for $z_{\rm s} = 5$. The predicted scattering is consistent with the expected reduction in $\Gscatt$, and is dominated by the contributions of dwarf galaxies, which outnumber the other galaxy types due to the GSMF.

\begin{figure}
    \centering
    \includegraphics[width=\linewidth]{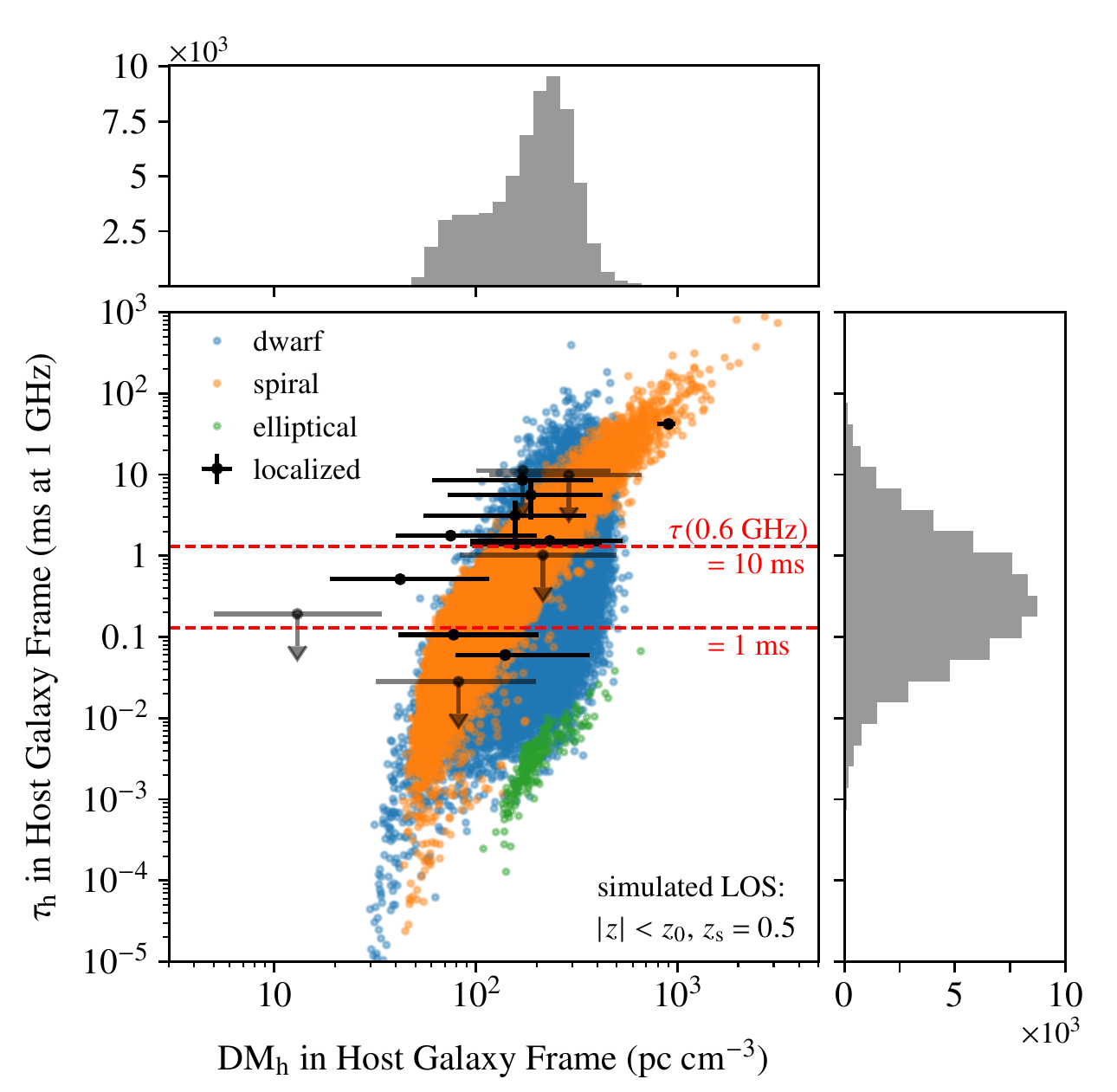}
    \caption{Host galaxy contributions of $\tau_{\rm h}$ vs. $\DMh$ for $10^4$ FRB LOS through dwarf (blue), spiral (orange), and elliptical (green) galaxies. Galaxy types were drawn using the galaxy stellar mass function from \cite{2021MNRAS.503.4413M}. Simulated results are shown in the galaxy rest frame for FRB locations within one density scale height of their host galaxy center and $\Ftilde$ referenced to $z_{\rm s} = 0.5$. The black points and error bars show estimated $\tau_{\rm h}$ vs. $\DMh$ for localized FRBs with scattering measurements, from the analysis in \cite{2021arXiv210801172C}. Scattering time upper limits are indicated by the black arrows. The grey histograms show the simulated distributions. The red dashed lines indicate $\tau = 10$ ms and $\tau = 1$ ms at 0.6 GHz scaled to 1 GHz.}
    \label{fig:frbsim_obs_compare}
\end{figure}

Expanding the distribution of FRB locations within their host galaxies significantly broadens the range of scattering times that may be expected. One extreme example is an older FRB progenitor that has migrated away from a galaxy disk, and is observed from the near edge of the galaxy. In this case, negligible scattering ($\tau\lesssim$ ns at 1 GHz) may be observed. The opposite extreme is an older FRB progenitor that has migrated towards the far edge of the galaxy, and is viewed through the entire galaxy disk. In this case, the scattering may be extremely large ($\tau \gg 100$ ms at 1 GHz), not only because the LOS samples a large fraction of the host galaxy ISM, but also because $\Gscatt>1$ when the FRB is offset from the scattering layer (which may also apply if, e.g., the FRB lies in a globular cluster in the halo). Even when the FRB locations are restricted to lie closer to their galaxy centers, the range of expected scattering times covers many orders of magnitude, which may suggest that scattering is a poor tool for distinguishing between different progenitor populations. 

The simulated distribution of $\tau$ vs. DM for host galaxies is compared to localized FRBs in Figure~\ref{fig:frbsim_obs_compare}. The estimated DM and scattering contributions of localized host galaxies are taken from \cite{2021arXiv210801172C}. The range of simulated DM and $\tau$ appears to be broadly consistent with the range of DM and $\tau$ constrained for localized FRBs, although a number of the localized sources have $\tau$ greater than the median of the simulated distribution. These results suggest that our estimates of $\tau$ may be conservative compared to the scattering observed from localized FRBs, \edit1{and may even be more conservative compared to the true population if selection effects bias the localized sample towards lower scattering \citep{seebeck2021}}.  Figure~\ref{fig:frbsim_obs_compare} also indicates that roughly half of the host galaxy $\tau$ distribution is greater than 1 ms at 0.6 GHz and should be measurable. 

\section{Cosmological Scattering Horizons}\label{sec:comp}

\begin{figure}
    \centering
    \includegraphics[width=\linewidth]{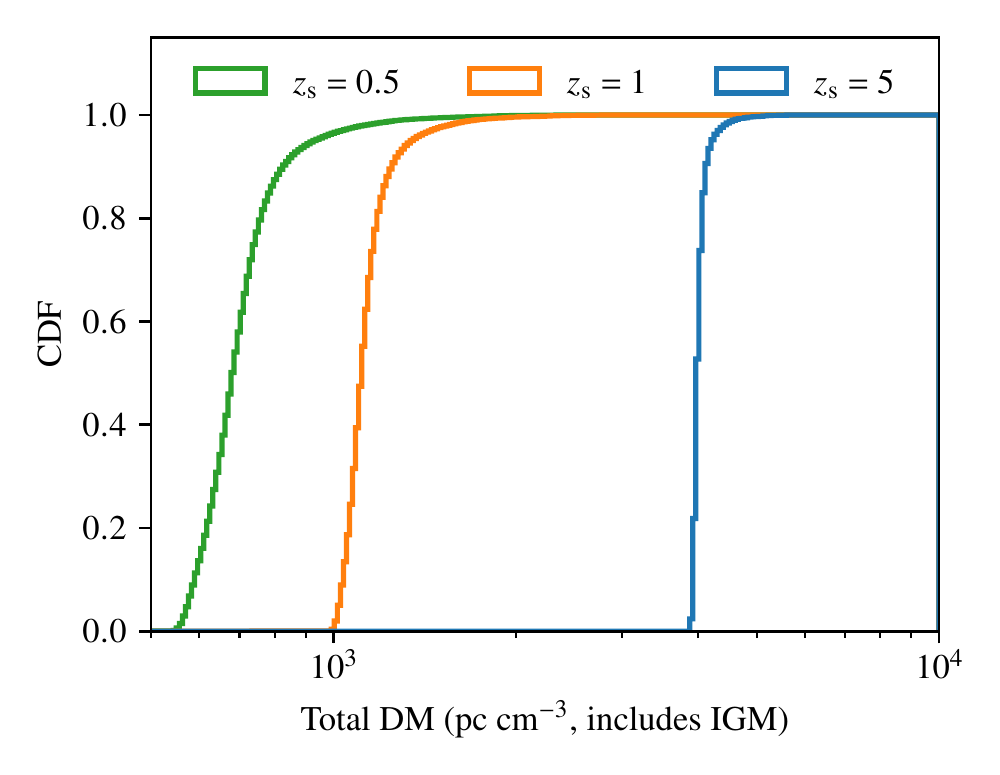}
    \caption{Cumulative distribution functions (CDFs) of the total simulated DM, including the IGM and all other LOS components considered in this study, in pc cm$^{-3}$ in the observer frame for $z_{\rm s} = 0.5$, $z_{\rm s} = 1$, and $z_{\rm s} = 5$. The distributions shown are equivalent to the sum of the DM components shown in Figure~\ref{fig:final_hist}, plus mean values for the IGM's DM contribution evaluated using Equation~\ref{eq:dmigm} at each source redshift: ${\rm \overline{DM}_{IGM}}(z_{\rm s} = 0.5) = 424$ pc cm$^{-3}$, ${\rm \overline{DM}_{IGM}}(z_{\rm s} = 1) = 870$ pc cm$^{-3}$, and ${\rm \overline{DM}_{IGM}}(z_{\rm s} = 5) = 3670$ pc cm$^{-3}$. Here we assume that variance in $\overline{\rm DM}_{\rm IGM}$ is induced by intersections through foreground halos, which are modeled independently. Additional variance in $\overline{\rm DM}_{\rm IGM}$ is possible if FRBs lie in overdense regions of the Universe \citep{2019ApJ...886..135P}, but this effect is not included here. }
    \label{fig:dm_with_igm}
\end{figure}

The \edit1{sum of} DM and scattering from the Milky Way ISM, nearby intervening galaxies in GWGC, distant intervening galaxies, and host galaxies are shown in the bottom panel of Figure~\ref{fig:final_hist} and in Table~\ref{tab:final_tab}. \edit1{Figure~\ref{fig:dm_with_igm} shows the total DM distribution, including the mean contribution of the IGM. Equation~\ref{eq:dmigm} is used to evaluate $\overline{\rm DM}_{\rm IGM}$ for each source redshift, yielding ${\rm \overline{DM}_{IGM}}(z_{\rm s} = 0.5) = 424$ pc cm$^{-3}$, ${\rm \overline{DM}_{IGM}}(z_{\rm s} = 1) = 870$ pc cm$^{-3}$, and ${\rm \overline{DM}_{IGM}}(z_{\rm s} = 5) = 3670$ pc cm$^{-3}$. Halo intersections are assumed to dominate deviations from $\overline{\rm DM}_{\rm IGM}$ and are modeled separately from the IGM component, although this does not include possible variance related to FRBs lying in overdense regions of the IGM \citep{2019ApJ...886..135P}. Inclusion of $\rm \overline{DM}_{\rm IGM}$ introduces a much stronger redshift evolution in the total DM distributions than is seen in any of the other individual DM components shown in Figure~\ref{fig:final_hist}.} \edit1{The median and $90\%$ c.i. of the total DM (observer frame) are $685^{+250}_{-107}$ pc cm$^{-3}$ for $z_{\rm s} = 0.5$, $1118^{+236}_{-91}$ pc cm$^{-3}$ for $z_{\rm s} = 1$, and $4010_{-75}^{+237}$ pc cm$^{-3}$ for $z_{\rm s} = 5$.}

The median and $90\%$ c.i. of $\tau$ (observer frame at 1 GHz) are $0.1_{-0.094}^{+2.3}$ ms for $z_{\rm s} = 0.5$, $0.1_{-0.093}^{+2.6}$ ms for $z_{\rm s} = 1$, and $0.08_{-0.07}^{+298}$ ms for $z_{\rm s} = 5$. For $z_{\rm s}\lesssim1$, host galaxies dominate the scattering distribution, whereas intervening galaxies dominate for $z_{\rm s} = 5$. About $20\%$ of the simulated FRBs from $z_{\rm s} = 5$ have $\tau > 5$ ms at 1 GHz. The cumulative scattering distributions for all three source redshifts overlap at $\tau \lesssim 0.1$ ms, largely because the host galaxy scattering for $z_{\rm s} = 0.5 - 1$ is comparable to the scattering from intervening halos for $z_{\rm s} = 5$ (although we note that these results are based on the host galaxy distribution where FRBs lie within about one ISM density scale height). As a result, re-scaling the distribution to 800 MHz yields about $40\%$ of the simulated FRBs with $\tau > 1$ ms, for all three source redshifts. 

\subsection{Comparison to CHIME/FRB Catalog 1}

\begin{figure}
    \centering
    \includegraphics[width=0.5\textwidth]{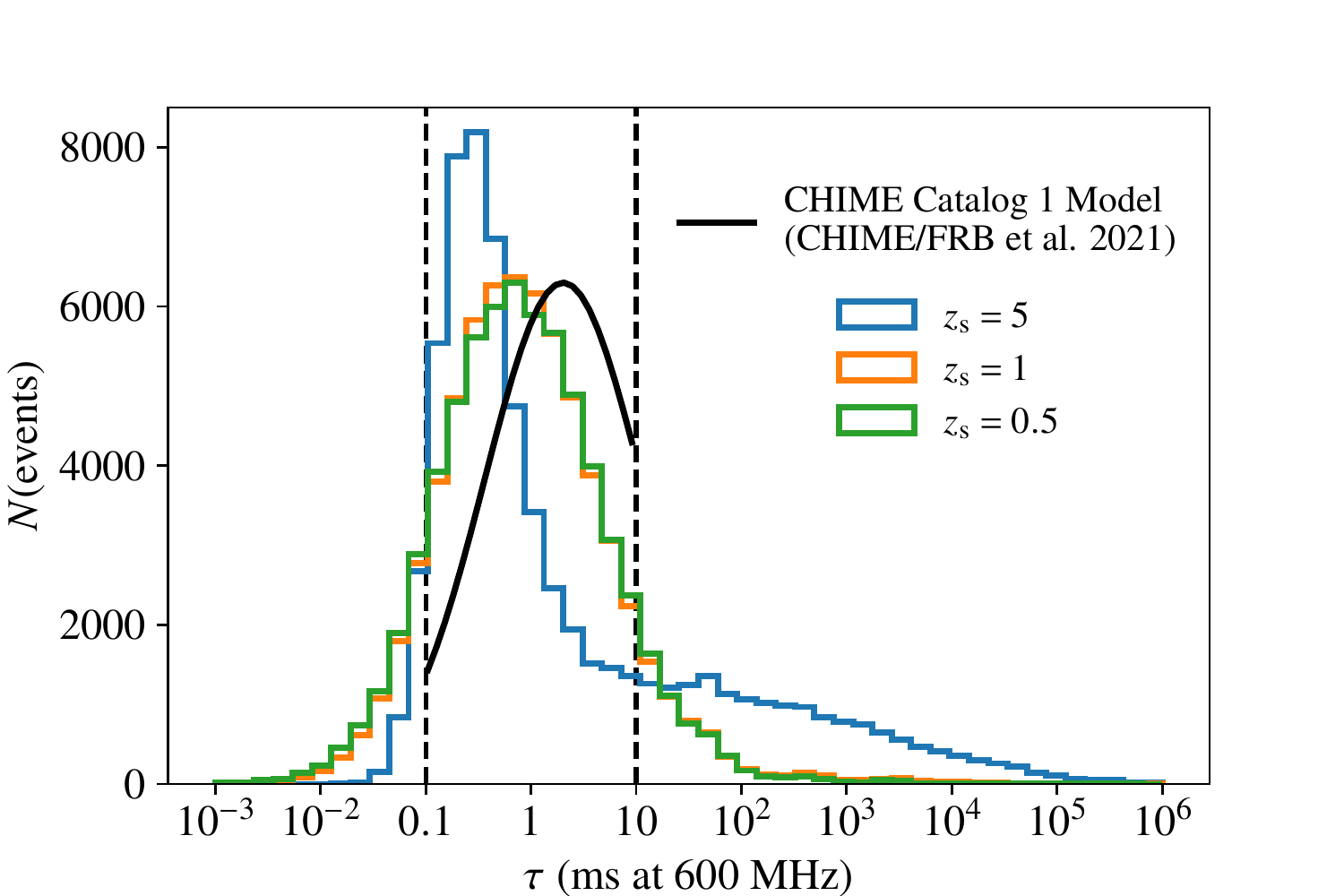}
    \caption{Histograms of the total scattering time distribution in ms at 600 MHz (re-scaled from 1 GHz assuming $\tau \propto \nu^{-4}$), including contributions from the Milky Way ISM, intervening galaxies, and host galaxies, for an all-sky population of $10^5$ FRBs. The scattering distribution was independently simulated for FRBs at three source redshifts: $z_{\rm s} = 0.5$ in green, $z_{\rm s} = 1$ in orange, and $z_{\rm s} = 5$ in blue. The black curve indicates the \cite{2021ApJS..257...59A} fiducial model for the intrinsic scattering distribution fit to CHIME/FRB Catalog 1 (including selection bias corrections), given in Appendix C of \cite{2021ApJS..257...59A} and re-scaled here to the number of simulated FRB events. The black dashed lines indicate the range of scattering times over which the CHIME distribution is constrained.}
    \label{fig:frbsim_chime_compare}
\end{figure}

\edit1{Figure~\ref{fig:frbsim_chime_compare} shows the simulated scattering time distributions compared to the \cite{2021ApJS..257...59A} fiducial model for the intrinsic scattering distribution fit to CHIME/FRB Catalog 1. The \cite{2021ApJS..257...59A} model accounts for selection effects that are constrained by injecting simulated bursts into the CHIME pipeline, and is fit to Catalog 1 using an iterative procedure that assumes scattering is uncorrelated with all other burst properties (namely fluence, DM, and burst width). The CHIME model is only constrained for $0.1 < \tau(600\ {\rm MHz}) < 10$ ms due to sensitivity limits. About half of the FRBs in Catalog 1 are estimated to be at redshifts $z_{\rm s}<0.5$ \citep{2021ApJS..257...59A}. The CHIME model is also fit to a measurement sample that includes a substantial number of FRBs with $\tau$ upper limits, and the model depends on the mitigation of a variety of systematics both instrumental (e.g., flux calibration) and observational (e.g., shape of the flux density spectrum). \cite{2021ApJS..257...59A} note that this model is predominantly meant to characterize selection effects in the catalog and is likely a rough estimate of the true, underlying distribution. We therefore refrain from a detailed statistical comparison to our simulated scattering distributions, and instead comment on the most obvious similarities and differences.}

\edit1{While the bulk of our simulated FRBs at $z_{\rm s} = 0.5$ and $z_{\rm s} = 1$ lie within the range of scattering times constrained by CHIME, the CHIME model peaks at  $\tau \approx 2$ ms at 600 MHz, about 1 ms larger than the peaks of our simulated distributions. The offset between the CHIME model and our simulated distributions is unsurprising, given that our simulations do not explicitly model the CHIME sample and do not incorporate any of the corresponding relevant constraints (e.g. estimated redshift distributions, energies, sky coverage, etc.). Explicitly considering the redshift distribution of CHIME sources would at least partially reconcile this difference because about half of Catalog 1 sources are estimated to be at $z_{\rm h} < 0.5$ \citep{2021ApJS..257...59A} and we find that lower redshift FRBs have scattering dominated by host galaxies. As a result, the amount of scattering from host galaxies at $z_{\rm h} < 0.5$ will be larger than the scattering that we simulate and show in Figure~\ref{fig:frbsim_chime_compare} for $z_{\rm s} = 0.5$ by a factor $[1.5/(1+z_{\rm h})]^3$. However, we note that the redshift distribution of CHIME sources is based on DM budgets that may also contain systematic biases. }

\edit1{Our simulations also indicate that a substantial fraction of highly scattered FRBs unobserved by CHIME may come from higher redshifts. This finding complements the \cite{2021arXiv210710858C} analysis of CHIME scattering, which used a different density modeling approach to argue that the scattering distribution observed by CHIME might require scattering from host galaxies or intervening halos that is enhanced compared to scattering considered typical of the Milky Way. While \cite{2021arXiv210710858C} predominantly considered scattering of lower redshift FRBs in an effort to reproduce the observed CHIME distribution, our simulations indicate that high-redshift ($z_{\rm s}>1$) FRBs could contribute to the regime of large scattering unconstrained by CHIME.}

\section{Summary \& Discussion}\label{sec:discussion}

We have modeled the dispersion and scattering of Galactic and extragalactic fast radio transients using a combination of NE2001 for the Milky Way \edit1{disk} and separate electron density models for \edit1{the Milky Way halo and} other galaxies that account for a range of galaxy types, masses, plasma densities, and strengths of turbulence. The chief results are summarized as follows:

\begin{itemize}
    \item Using NE2001, we provide a latitudinal and frequency-dependent prescription for the Galactic scattering zone of avoidance, which is largely confined to $|b| < 5^\circ$ and severely impacts the detection of pulsar and FRB LOSs near the inner Galaxy in the time domain. For an all-sky population of FRBs, Galactic scattering contributes negligibly to the total scattering of the population. 
    \item The range of DM contributions predicted for nearby ($\dlo \lesssim 100$ Mpc) galaxies is broadly consistent with those found in other studies \citep[e.g.][]{2019MNRAS.485..648P, 2021arXiv210713692C}, whereas the predicted scattering contributions of these galaxies for an all-sky population of FRBs is extremely small ($90\%$ with $\tau<0.02$ ms at 1 GHz). Pulsars and FRBs residing in or intersecting the Magellanic Clouds will provide critical constraints on the turbulent fluctuation parameter $\Ftilde$ in these satellite galaxies, which may substantially scatter radio transients.
    \item Most FRBs at redshifts $z_{\rm s} \gtrsim 0.3$ will be seen within twice the virial radius of at least one halo with mass $\lesssim 10^{14}M_\odot$. Scattering from a distribution of galaxies intervening an FRB LOS is predicted to be $\lesssim 0.01$ ms at 1 GHz ($90\%$ confidence) for $z_{\rm s} \leq 1$, whereas as many as $20\%$ of higher redshift FRBs ($z_{\rm s} \sim 5$) may have $\tau > 1$ ms at 1 GHz from intervening galaxies alone.
    \item Host galaxies are predicted to dominate the scattering budgets of FRBs at $z_{\rm s} \leq 1$, with $\tau \lesssim 2$ ms at 1 GHz ($90\%$ confidence; this value does not include enhancements to near-source environments). The estimated DM distribution for host galaxies (median $\approx 200$ pc cm$^{-3}$, galaxy frame) is broadly consistent with those estimated in recent studies \citep[e.g.][]{2021arXiv210801172C,2021arXiv210710858C,2022MNRAS.509.4775J} and favors larger DMs than assumed in some previous studies \citep[e.g.][]{2020Natur.581..391M,2021AA...651A..63G}. 
    \item A cumulative assessment of scattering from host galaxies, intervening galaxies, and the Milky Way indicates that over $40\%$ of FRBs from redshifts $z_{\rm s} \geq 0.5$ may have $\tau \gtrsim 1$ ms at frequencies $\nu \leq 800$ MHz. We find that $20\%$ of a high-redshift $(z_{\rm s} = 5)$ FRB population will have $\tau > 5$ ms at 1 GHz. We therefore find a substantial fraction of FRBs with large scattering, despite a fairly conservative electron density model that is homogeneous and ignores local structures that can further amplify scattering. 
\end{itemize}

\noindent These results are based on a few key assumptions about the relationship between FRB progenitors and host galaxies, the connection between electron density and a galaxy's optical morphology, and the properties of circumgalactic turbulence. These assumptions and suggestions for future work are discussed below.

\subsection{Host Galaxies \& FRB Progenitors}

\indent The estimated host galaxy contributions to DM and scattering depend on the relationship between FRB progenitors and their hosts. In this study we are agnostic about physical sources of FRBs, and are primarily concerned with scattering from the host galaxy ISM. The results listed above are for FRBs located within one density scale height of their ISM, but as we discuss in Section~\ref{sec:hosts}, the range of scattering from hosts can be much larger if the distribution of FRBs within host galaxies is expanded to a broader range of progenitor locations. In addition, our density modeling does not include plasma structure locally related to an FRB source. Distinguishing between ISM and near-source scattering is possible with measurements of pulse broadening and Galactic diffractive interstellar scintillation, which can jointly constrain the distance between the FRB source and the dominant extragalactic scattering material \citep[e.g.][]{2015Natur.528..523M, 2022arXiv220213458O}. 

\indent \edit1{Circumsource environments can only contribute significantly to observed scattering under specific conditions that depend on the density of the surrounding material (and thus whether free-free absorption affects burst propagation) and the distance between the source and circumsource plasma. If the circumsource material is confined to a certain radius, the scattering time can be no larger than the propagation time across that radius and will not scale according to the usual $\nu^{-4}$ due to the finite size of the scattering region (the ``truncated screen" effect; \citealt{2001ApJ...549..997C}). For example, the maximum delay from a shell of radius $r$ around a source at $\dso = 1$ Gpc would be $\tau_{\rm max} \sim r^2 / (2c \dso) \sim 50\ {\rm ms} \times (r / 1\ {\rm pc})^2 (\dso/1\ {\rm Gpc})^{-1}$. In practice, the scattering delay could be much smaller than 50 ms at 1 GHz and will not scale as $\nu^{-4}$. The relevant size scales of circumsource plasma is also highly uncertain, and could range anywhere from $\sim10$s au to $10$s of pc \citep[e.g.][]{2022arXiv220211112A,2022arXiv220213458O}. It is thus unclear to what extent circumsource environments could contribute to scattering of the FRB population.} 

\indent Regardless of the exact scattering geometry within a host galaxy, the corresponding scattering horizon will largely be confined to the inner galaxy, similar to the Milky Way. As such, FRBs localized to apparently small offsets from their host galaxy centers may lie in the foreground of the inner galaxy if the FRB appears to be unscattered. Scattering horizons within hosts may complicate the interpretation of FRB localizations in terms of progenitor populations. Most if not all FRBs localized to spiral arm galaxies are located on the outer arms, and it has been argued that the population of known, localized FRBs does not support progenitor channels involving massive stars or neutron star mergers \citep{2021ApJ...917...75M}. Future comparisons of the FRB population with different progenitor models should consider whether scattering within host galaxies biases the apparent distribution of FRBs within their hosts, \edit1{ as has been noted by, e.g., \cite{seebeck2021}}. Even if VLBI localizations are typically performed at higher frequencies where scattering is minimized, detection bias from scattering will affect which initial FRB detections trigger VLBI follow-up. Surveys that perform detection and localization simultaneously (such as ASKAP and future projects like CHIME outriggers, DSA-2000, and CHORD) may therefore reduce this bias. 

\subsection{Connection between Electron Density, Turbulence, \& Other Galaxy Properties}

We have constructed density models for three galaxy types, dwarfs, spirals, and ellipticals, in order to estimate the range of dispersion and scattering that may be expected from different galaxy morphologies. The plasma distributions and turbulence of these galaxy types were modeled based on a combination of \edit1{pulsar and FRB observations probing the Milky Way and Magellanic Clouds, and} the relative strengths of H$\alpha$ emission and SFR seen in galaxy surveys, as well as their typical stellar masses. While we have adopted PDFs for the central density and $\Ftilde$ meant to approximate the breadth of properties seen even among galaxies of the same type, \edit1{the predicted range of scattering times could increase substantially if the distribution of $\Ftilde$ for the galaxy population is even broader than assumed in this study.} Future work may benefit from a more direct consideration of the relationships between plasma density and physical processes that operate on scales smaller than optical morphology, such as the relationship between gas density and SFR (i.e. the Kennicutt-Schmidt law, \citealt{2012ARAA..50..531K}). Future modeling may also want to tie electron density scale lengths to stellar mass, rather than to halo mass as we have done, because early and late-type galaxies show distinct relationships between their radii and stellar masses that also evolve with redshift \citep{2014ApJ...788...28V}.

We have also assumed that the fluctuation parameter $\Ftilde$ broadly traces SFR$(z)$, as the ionized gas velocity dispersion in nearby disk galaxies is consistent with turbulence driven by star formation \citep{2010Natur.467..684G,2021arXiv211211281L}. However, the lack of comparably large samples at higher redshifts makes it unclear whether other processes, such as gravitational instabilities, drive turbulence beyond the peak of the CSFR \citep{2017MNRAS.467.4080F, 2018MNRAS.477...18P}. Unlike other spectroscopic tracers of ionized gas, radio wave propagation is insensitive to ionized gas temperature, and probes turbulent spatial fluctuations rather than turbulent velocities. Continued analysis of FRBs localized to host galaxies will provide critical, independent means of connecting electron density profiles to FRB-specific observables, which will help circumvent the uncertainties related to converting between different ISM tracers. 

\subsection{Circumgalactic Turbulence}

The distribution of $\Ftilde$ adopted for the CGM was based predominantly on scattering observations of localized FRBs intersecting the Milky Way halo and the halos of M33 and M81. This distribution is broadly consistent with CGM turbulence that is substantially weaker than turbulence in the warm ionized ISM, and may be consistent with expectations for a CGM dominated by very diffuse, hot ionized gas \citep{2021ApJ...911..102O}. While there is substantial evidence that the CGM can be a multi-phase medium \citep{2016ApJS..226...25L,2018MNRAS.473.5407M,2020MNRAS.491.5056L}, it remains unclear whether cooler, clumpy CGM gas meaningfully contributes to FRB scattering \edit1{via density fluctuations on the $\sim$sub-au scales necessary to produce diffractive, multi-path propagation}. 

\cite{2019MNRAS.483..971V} model radio wave scattering from cooler CGM cloudlets, which are estimated to have a large areal covering fraction but small volume filling factor \citep{2020MNRAS.491.5056L}. While \cite{2019MNRAS.483..971V} find that radio wave scattering from cooler CGM gas may only be relevant in a limited halo mass range between about $10^{11.5}$ and $10^{13.5}M_\odot$, they estimate a relatively large amount of scattering from these cool cloudlets, $\tau \gtrsim 1$ ms at 1 GHz for $z_{\rm s} \geq 1$. The scattering predicted by this model for low-$z$ halos is much larger than the scattering observed from FRBs intersecting the halos of the Milky Way and nearby galaxies, but redshift evolution in, e.g., the turbulence outer scale and gas filling factor may have a strong effect on the scattering observed from higher redshift halos.

\subsection{Galaxy Clusters}

Throughout this study we have omitted special modeling of galaxy clusters, instead treating clusters as single, large-mass halos. While cluster intersections are less common than intersections with lower-mass halos, future modeling should examine the possibility of enhanced density and turbulence in the intracluster medium. Moreover, the fluctuation parameter may be different in the outer and inner regions of the cluster, due to the relative mediation of turbulence by IGM gas infall (which may produce accretion shocks) and AGN feedback \citep[e.g.][]{2021MNRAS.508.1777B, 2021ApJ...920..104P}. FRB observations of clusters may be able to resolve these variations in $\Ftilde$ if a large sample of FRBs is used to infer cluster DM and $\tau$ profiles, and if the cluster and FRB redshifts are known (which breaks the degeneracy between $\Ftilde$ and $\Gscatt$). In theory, a sample of FRB LOS through a galaxy cluster has the potential to self-consistently probe the intracluster medium across both the inner regions typically constrained with X-ray observations and the outer regions constrained by observations of the Sunyaev-Zeldovich effect. 

\section{Conclusions}\label{sec:conclusions}

We have characterized a fiducial amount of scattering from the Milky Way, intervening galaxies, and host galaxies along pulsar and FRB LOSs. For a median burst width $W \approx 1$ ms at 600 MHz (based on CHIME/FRB Catalog 1 and ignoring selection effects; \citealt{2021ApJS..257...59A}), about $45\%$ of FRBs at redshifts $z_{\rm s} \leq 1$ and $40\%$ of FRBs at redshift $z_{\rm s} \sim 5$ will have $\tau \geq W$ and are selected against due to scattering horizons alone. These percentages could be lower limits given that the amount of scattering predicted by our electron density modeling may be conservative when compared to \edit1{the current sample of} localized FRBs, \edit1{which may already be biased towards low scattering}. \edit1{The extent to which the FRB population is affected by scattering depends primarily on how often FRBs encounter regions of high scattering.} Circumgalactic turbulence with a larger $\Ftilde$ and the inclusion of discrete structures \edit1{within galaxy ISMs (and possibly near FRB sources) could therefore} increase the \edit1{predicted} impact of scattering \edit1{on the FRB population}. Upcoming surveys performed with, e.g., CHIME outriggers and DSA-2000 may provide 1000s of localized FRBs over the next decade \citep{2019BAAS...51g.255H, 2021AJ....161...81L}. These localizations will enable the construction of precise scattering budgets that will constrain the redshift evolution of $\Ftilde$, and allow us to statistically resolve the plasma densities of galaxies along FRB LOSs as functions of their stellar and halo masses, sizes, star formation rates, and strengths of turbulence. 

\acknowledgements{The authors thank Prof. Nick Battaglia, Prof. Martha Haynes, and Catie Ball for illuminating conversations that contributed directly to this work, and thank the anonymous referee for their comments. S.K.O., J.M.C., and S.C. acknowledge support from the National Aeronautics and Space Administration (NASA 80NSSC20K0784). The authors also acknowledge support from the National Science Foundation (NSF AAG-1815242) and are members of the NANOGrav Physics Frontiers Center, which is supported by the NSF award PHY-2020265. M.G. was supported by the Research Experiences for Undergraduates at Cornell University. Cornell is located on the traditional homelands of the Gayog\b{o}h\'{o}:n\c{o}\'{} (the Cayuga Nation), members of the Haudenosaunee Confederacy that precedes the establishment of Cornell University, New York State, and the United States of America.}

\bibliography{bib}

\listofchanges

\end{document}